\documentclass{elsart}
\usepackage{graphics}
\usepackage{amssymb}

\begin{document}

\begin{frontmatter}

\title{Statistical mechanics methods and \\phase transitions
in optimization  problems.}

\author[label1]{Olivier C. Martin\thanksref{email1},}
\author[label2]{R\'emi Monasson\thanksref{email2},}
\author[label3]{and Riccardo Zecchina\thanksref{email3}}

\address[label1]{LPTMS, Universit\'e Paris-Sud, Orsay, France}
\address[label2]{The James Franck Institute, The University of Chicago, 
Chicago, Il.}
\address[label3]{International Centre for Theoretical Physics, Trieste, Italy}

\thanks[email1]{E-mail: martino@ipno.in2p3.fr}
\thanks[email2]{Permanent address: 
CNRS-Laboratoire de Physique Th\'eorique de l'ENS, Paris, France; 
E-mail: monasson@lpt.ens.fr}
\thanks[email3]{E-mail: zecchina@ictp.trieste.it}
\begin{abstract}
Recently, it has been recognized that phase transitions play an
important role in the probabilistic analysis of combinatorial
optimization problems. However, there are in fact many other relations
that lead to close ties between computer science and statistical
physics. This review aims at presenting the tools and concepts
designed by physicists to deal with optimization or decision problems
in an accessible language for computer scientists and mathematicians,
with no prerequisites in physics. We first introduce some elementary
methods of statistical mechanics and then progressively cover the
tools appropriate for disordered systems. In each case, we apply these
methods to study the phase transitions or the statistical properties
of the optimal solutions in various combinatorial problems. We cover
in detail the Random Graph, the Satisfiability, and the Traveling
Salesman problems. References to the physics literature on
optimization are provided. We also give our perspective regarding the
interdisciplinary contribution of physics to computer science.
\end{abstract}

\end{frontmatter}

\newpage

\section{Introduction}
\label{sect_intro}

At the heart of statistical physics, discrete mathematics, and
theoretical computer science, lie mathematically similar counting and
optimization problems.  This situation leads to a transgression of
boundaries so that progress in one discipline can benefit the others.
An old example of this is the work of Kasteleyn (a physicist) who
introduced a method for counting perfect matchings over planar graphs
(a discrete mathematics problem). Our belief is that a similar
cross-fertilization of methods and models should arise in the study of
combinatorial problems over {\it random structures}. Such problems
have attracted the attention of a large community of researcher in the
last decade, but a transgression of boundaries has only just
begun. One of the many potential spin-offs of this kind of
cross-fertilization would be the use of computer science and graph
theoretical methods to tackle unsolved problems in the statistical
physics of ``complex'' (disordered) systems. But we also hope that the
benefits can go the other way, {\it i.e.}, that the recent
developments in statistical physics may be of use to the other two
communities; such is our motivation for this article.

This review does not assume any knowledge in physics, and thus we
expect it to be accessible to mathematicians and computer scientists
eager to learn the main ideas and tools of statistical physics when
applied to random combinatorics.  We have chosen to illustrate these
``physical'' approaches on three problems: the Random Graph, the
Satisfiability, and the Traveling Salesman problems. This particular
focus should help the interested reader explore the statistical
physics literature on decision and optimization problems. Furthermore,
we hope to make the case that these methods, developed during the last
twenty years in the context of the so called spin glass
theory~\cite{MezardParisiVirasoro,Young98}, may provide new concepts
and results in the study of {\em phase transitions}, and {\it average}
case computational complexity, in computer science problems. Some
examples of this kind of methodological transfer can also be found in
three other papers of this TCS special issue, dealing with statistical
mechanics analyses of vertex covering on random
graphs~\cite{HartWeigt}, of number partitioning~\cite{Mertensrev} and
of learning theory in artificial neural networks~\cite{Engel}.

Random combinatorics became a central part of graph theory following
the pioneering work by Erd\"os and R\'enyi. Their study of clusters in
random graphs (percolation for physicists) showed the existence of
zero-one laws (phase transitions in the terminology of physics). More
recently, such phenomena have played a fundamental role when tackling
average--case complexity. Indeed, numerical evidence suggests that the
onset of intractability in {\it random} NP-complete problems can be
put in relation with the appearance of phase transitions analogous to
the percolation transition.  Interestingly, the concept of random
structures is present in most natural sciences, including biology,
chemistry, or physics. But in the last two decades, the theoretical
framework developed in physics has lead to new analytical and
numerical tools that can be shared with the more mathematical
disciplines.  The potential connections between discrete mathematics,
theoretical computer science and statistical physics become
particularly obvious when one considers the {\it typical} properties
of random systems. In such cases, percolation, zero-one laws, or phase
transitions are simply different names describing the same phenomena
within the different disciplines. It seems to us that much can be
gained by exploring the complementary nature of the different
paradigms in mathematics and physics.  In what follows, we shall try
to make this happen by giving a thorough statistical mechanics
analysis of three prototype problems, namely percolation in random
graphs, satisfiability in random K-Satisfiability, and optimization
via the Traveling Salesman Problem. The review is preceded by a
general discussion of some basic concepts and tools of statistical
mechanics. We have also included simple exercises to help the
interested reader become familiar with the methodology; hopefully he
(she) will be able to adapt it to the study of many other problems,
e.g., matching, number partitioning \cite{Mertensrev}, etc...  When
appropriate, we compare the results of statistical physics to those of
discrete mathematics and computer science.

From a statistical mechanics perspective, a phase transition is
nothing but the onset of non-trivial macroscopic (collective) behavior
in a system composed of a large number of ``elements'' that follow
simple microscopic laws. The analogy with random graphs is
straightforward.  There the elements are the edges of the graph which
are added at random at each time step and the macroscopic phenomenon
is the appearance of a connected component of the graph containing a
finite fraction of all the vertices, in the limit of a very large
number of vertices.  If a system has a phase transition, it can be in
one of several ``phases'', depending on the values of some control
parameters. Each phase is characterized by a different microscopic
organization. Central to this characterization is the identification
of an {\it order parameter} (usually the expectation value of a
microscopic quantity) which discriminates between the different
phases. Once again the analogy with random graphs is appropriate. An
order parameter of the percolation transition is the fraction of
vertices belonging to the giant connected component. Such a fraction
is zero below the percolation transition, that is, when the
connectivity of the random graph is too small, and becomes strictly
positive beyond the percolation threshold.

While in percolation it is proven that the order parameter is indeed
the fraction of vertices belonging to the infinite giant component, in
more complicated systems the determination of an order parameter is
generally an open problem.  Though not rigourous, statistical
mechanics provides numerous specific methods for identifying and
studying order parameters, and we shall illustrate this on the
K-Satisfiability problem. This step is useful of course for providing
a good intuitive view of the system's behavior, but more importantly
it also gives information on the microscopic structure of the phases,
information that can be used both in deriving analytical results and
in interpreting numerical simulations.

The way physicists and mathematicians proceed is quite different.
Theoretical physicists generally do not prove theorems, rather they
attempt to understand problems by obtaining exact and approximate
results based on reasonable hypotheses. In practice, these hypotheses
are ``validated'' {\em a posteriori} through comparison with
experiments or numerical simulations, and through consistency with the
overall body of knowledge in physics.  In this sense, theoretical
physics must be distinguished from mathematical physics whose scope is
to make rigorous statements.  Of course, exact solutions play an
important role in statistical physics in that they represent limiting
cases where analytical or numerical techniques can be checked, but
they are not the main focus of this discipline.

For the sake of brevity we left out from this review some very
relevant and closely connected topics such as exact enumeration
methods~\cite{baxter} or applications of computer science algorithms
to the study of two dimensional complex physical systems
\cite{Rieger98,Youngrev}. Furthermore we do not claim to present
a {\it complete} picture of what has been done by physicists on
decision and optimization problems. Rather, we hope that what we do
present will enable readers from the more mathematical disciplines to
understand in detail the {\it majority} of what has been done by
physicists using the methods of statistical mechanics.

\section{Elements of Statistical Physics} \label{sect_statisticalPhysics}

In this section, the reader will be introduced to the basic notions of
statistical mechanics. We start by illustrating on various examples
the existence of 
phases and phase transitions, ubiquitous in physics and more surprisingly 
in other fields of science too. The concepts of 
microscopic and macroscopic levels  of description naturally
appear and allow for a rapid presentation of the foundations of statistical 
mechanics. We then expose in greater detail the combinatorial
interpretation of statistical mechanics and introduce some key vocabulary and
definitions. An accurate investigation of the properties of the so-called
Ising model on the complete graph $K_N$ exemplifies the above concepts
and calculation techniques. In order to bridge the gap with 
optimization problems, we then turn to the crucial issue of randomness
and present appropriate analytical techniques to deal with random structures,
e.g., the celebrated replica method.

This section has been elaborated for a non physicist readers and we stress 
that no {\em a priori} knowledge of statistical mechanics is required. 
Exercises have been included to illustrate key notions and
should help the reader to acquire a deeper understanding of concepts and
techniques. Solutions are sketched in Appendix A. 
Excellent presentations of statistical mechanics can be found in textbooks
e.g.\cite{Reif,Ma,Huang} for readers wanting further details.

\subsection{Phases and transitions}

Many physical compounds can exist in nature as distinct ``states'',
called phases, depending on the values of control parameters,
such as temperature, pressure, ... The change of phase happens very
abruptly at some precise values of the parameters and is called
transition. We list below a few well-known examples from
condensed matter physics as well as two cases
coming from biology and computer science.

\subsubsection{Liquid-gas transition.}

At atmospheric pressure water boils at a ``critical'' temperature
$T_c=100^o$C. When the temperature $T$ is lower than $T_c$, water is
a liquid while above $T_c$ it is a gas. At
the critical temperature $T_c$, a coexistence between the liquid and
gas phases is possible: the fraction of liquid water
depends only on the total volume occupied by both phases. 
The coexistence of the two phases at criticality is
an essential feature of the liquid-gas transition. Transitions sharing
this property are called {\em first order} phase transitions for
mathematical reasons exposed later.

\subsubsection{Ferromagnetic-paramagnetic transition.}

It is well-known that magnets attract nails made out of iron. The
magnetic field produced by the magnet induces some strong internal
magnetization in the nail resulting in an attractive force. 
Materials behaving as iron are referred to as
ferromagnetic. However, the attractive force disappears when the
temperature of the nail is raised above $T_c=770^oC$. The nail then
enters the paramagnetic phase where the net magnetization
vanishes.
There is no phase coexistence at the critical temperature; the
transition is said to be of {\em second order}.

The ferromagnetic-paramagnetic transition temperature $T_c$ varies
considerably with the material under consideration. For instance,
$T_c=1115^o$C for cobalt, $T_c=454^o$C for nickel and $T_c=585^o$C for
magnetite (Fe$_3$O$_4$). However, remarkably, it turns out that some other
quantities -- the critical exponents related to the (drastic) changes
of physical properties at or close to the transition -- are
equal for a large class of materials! The discovery of such
{\em universality} was a breakthrough and led to very deep
theoretical developments in modern physics. Universality is
characteristic of second order phase transitions.

\subsubsection{Conductor-superconductor transition.}

Good conductors such as copper are 
used to make electric wires because of their weak 
resistance to electric currents at room temperature. As the temperature
is lowered, electrical resistance generally decreases smoothly as
collisions between electrons and vibrations of the metallic crystal become
weaker and weaker. In 1911, Kammerling Onnes observed that the electrical
resistance of a sample of mercury fell abruptly down to zero as temperature
passed through $T_c\simeq 4.2^oK$ ($0^oK$ being the absolute
zero of the Kelvin scale.) This change of state, between a normal
conductor (finite resistance) and a superconductor (zero resistance)
is a true phase transition: a very small variation of temperature at $T_c$
is enough to change resistance by four or five orders of
magnitude!

\subsubsection{DNA denaturation transition.}

In physiological conditions, DNA has the double helix structure
discovered by Watson and Crick in 1953. The two strands carry
complementary sequences of A, T, G or C bases and are intertwined, forming
either A-T or G-C pairs. Bases in a pair are attached
together by hydrogen bonds. As the temperature is raised or ionic
conditions are appropriately modified, bonds weaken and break
up. The strands may then separate so that the double helix structure is lost:
the DNA is denatured. This transition is abrupt on repeated
homogeneous DNA sequences~\cite{Saenger84}.

Recent micromanipulation experiments on individual DNA molecules have
shown that denaturation can also be obtained through a mechanical
action on DNA. When imposing a sufficient torque to the molecule to
unwind the double helix, the latter opens up and DNA denaturates. At a
fixed critical torque, denaturated and double helix regions may
coexist along the same molecule\cite{Strick98} so this transition is like
a liquid-gas one.

\subsubsection{Transition in the random K-Satisfiability problem.}

Computer scientists discovered some years ago that the random
K-Satisfiability problem exhibits a threshold phenomenon as the ratio
$\alpha$ of the number of clauses ($M$) over the number of Boolean
variables ($N$) crosses a critical value $\alpha _c (K)$ depending on
the number of literals per clause $K$. When $\alpha$ is
smaller than the threshold $\alpha _c (K)$, a randomly drawn formula
is almost surely satisfiable while, above threshold, it is
unsatisfiable with probability reaching one in the $N\to \infty$ limit.

For $K=2$, the threshold is known exactly: $\alpha _c
(2)=1$.  For $K \ge 3$, there is no rigorous proof of the existence of
a phase transition so far but many theoretical and numerical results
strongly support it, see articles by Achlioptas \& Franco and Dubois
\& Kirousis in the present issue.  Current best estimates indicate that the
threshold of random 3-SAT is located at $\alpha _c (3) \simeq 4.25$.
Statistical physics studies show that the order of the phase transition
depends on $K$, the transition being continuous for 2-SAT and of first
order for 3-SAT (and higher values of $K$).

\subsubsection{Macroscopic vs. microscopic descriptions.}

What can be inferred from the above examples? First, a (physical)
system may be found in totally different phases 
with very different macroscopic properties
although its intrinsic
composition at a microscopic level (molecules, magnetic spins, base pairs,
clauses, ...)  is the same. However, from a
physical, mechanical, electrical, biological, computational, ...
point of view, essential properties of this system change completely
from a phase to another. Second, the {\em abrupt} change of phase 
follows from very slight modifications of a control parameter
e.g. temperature, torque, ratio of clauses per variable ... about
a critical value. Thirdly, critical exponents, that characterize
quantitatively second order phase transitions, are universal, that is,
insensitive to many details of the systems under study. Last of all,
transitions appear for large systems only.

The above points raise some fundamental questions: how can the main
features of a system at a macroscopic level, defining a phase, change
abruptly and how are these features related to the microscopic
structure of the system?  Statistical physics focuses on these questions.

\subsection{Foundations of statistical mechanics and relationship with
combinatorics.}

\subsubsection{Needs for a statistical description.}

Statistical physics aims at predicting
quantitatively the macroscopic behaviour of a system (and in
particular its phases) from the
knowledge of its microscopic components and their interactions. 
What do we mean by interaction? Consider for instance a liquid made
of $N$ small particles (idealized representations of
atoms or molecules) occupying positions of coordinates $\vec r_i$ 
in Euclidean space
where label $i$ runs from 1 to $N$. Particle number $i$ is subject to
a force $\vec f_i$ (interaction) 
due to the presence of neighboring particles;
this force generally depends of the relative positions of these particles.
To determine the positions of the
particles at any later time
$t$, we must integrate the equations of motion given by Newton's 
fundamental law of mechanics,
\begin{equation}
m_i \; \frac{d^2 \vec r_i}{dt^2} = \vec f_i ( \{ \vec r_j \}) \ , \qquad
(i=1,\ldots, N) ,
\label{Newton}
\end{equation}
where $m_i$ is the mass of particle $i$. Solving these equations cannot be done
in practice. The forces $\vec f_i$ are indeed highly non 
linear functions of the particle positions $\vec r_j$. We therefore wind up
with a set of complicated coupled
differential equations whose number $N$, of order
$\sim 10^{23}$, is gigantic and not amenable to analytical treatment.

This impossibility, added to the intuitive feeling that understanding 
macroscopic properties cannot require the exact knowledge of all microscopic
trajectories of particles has been circumvented by a totally 
different approach.
The basic idea is to describe the system of particles in a probabilistic way
in order to deduce macroscopic features as emergent statistical properties.

\subsubsection{Probability distribution over the set of configurations.}

The implementation of this idea has required the introduction of 
revolutionary concepts at the end of the ninteenth century by Boltzmann and
followers, and in particular, the ideas of ergodicity and
thermodynamical equilibrium. We shall not attempt here to provide an 
exposition of these concepts. 
The interested reader can consult textbooks e.g. \cite{Reif,Ma,Huang}.
As far as combinatorial aspects of statistical mechanics are concerned, it is 
sufficient to start from the following postulate.

A configuration ${C}$  of the system, that is, the specification of the $N$
particle positions $\{ \vec r_j \}$, has a probability $p({C} )$ to be
realized at any time when the system is
in equilibrium. In other words, the system will be in configuration
${C}$ with probability $p({C})$. The latter depends on temperature
$T$ and equals
\begin{equation}
p(C ) = \frac 1Z \; \exp \left( -\frac 1T \; E({ C } )
\right) \quad .
\label{gibbs}
\end{equation}
In the above expression, $E$ is the {\em energy} and is a real-valued  
function,  over 
the set of configurations. The {\em partition function} $Z$ ensures the correct
normalization of the probability distribution $p$,
\begin{equation}
Z = \sum _C \exp \left( -\frac 1T \; E({ C } )
\right) \quad .
\label{foncpart}
\end{equation}
Note that we have used a discrete sum over configurations $C$ 
in (\ref{foncpart}) instead of an integral over particle 
positions $\vec r_j$. This notation has been chosen since all the partition 
functions we shall meet in the course of studying
optimization problems are related to finite ({\em i.e.} discrete) sets 
of configurations. 

Consider two limiting cases of (\ref{gibbs}):
\begin{itemize}
\item {\em infinite temperature $T=\infty$:}
the probability $p(C)$ becomes independent of $C$. All configurations
are thus equiprobable. The system is in a fully ``disordered'' phase,
like a gas or a paramagnet.
\item {\em zero temperature $T=0$:} 
the probability $p(C)$ is concentrated on the minimum of the energy
function $E$, called the {\em ground state}.  This minimum corresponds to
a configuration where all particles are at mechanically stable
positions, that is, occupy positions $r_i$ carefully optimized so that
all forces $f_i$ vanish. Often, these strong constraints define
regular packings of particles and the system achieves a perfect
crystalline and ``ordered'' state.
\end{itemize}
When varying the temperature, intermediate situations can be reached.  We
now examine some simple examples.

\subsubsection{Cases of one and two spins.}

We now consider the case of a single abstract particle that can sit at two
different positions only. This simple system can be recast as follows. 
Let us imagine an arrow capable of pointing in the up or down 
directions only. This arrow is usually called a {\em spin} and the direction 
is denoted by a binary variable $\sigma$, equal to $+1$ if the spin is up,
to $-1$ if the spin is down. 

In this single particle system,
there are only two possible configurations $C=\{ +1 \}$
and $C=\{ -1 \}$ and we choose for the energy function
$E(\sigma)= - \sigma$.
Note that additive constants in $E$ have no effect on (\ref{gibbs}) 
and multiplicative constants can be absorbed in the temperature $T$.
The partition function can be easily computed from
(\ref{foncpart}) and reads $Z=2 \cosh \beta$ where 
$\beta = 1/T$ denotes the inverse temperature. The probabilities that the spin 
points up or down are respectively $p_+ = \exp (
\beta )/Z$ and $p_- = \exp ( -\beta )/Z$. At infinite temperature ($\beta=0$), 
the spin is indifferently up or down: $p(+1)=p(-1)=1/2$. Conversely,
at zero temperature, it only points upwards: $p(+1)=1,p(-1)=0$.
$C=\{ +1 \}$ is the 
configuration of minimum energy. 

The average value of the spin, called magnetization is given by
\begin{equation}
m = \langle \sigma \rangle _T= \sum _{\sigma=\pm 1} \sigma
 \; p(\sigma) = \tanh (\beta ) \quad . \label{mag1}
\end{equation}
The symbol $\langle \cdot \rangle _T$ denotes the average over the probability 
distribution $p$. Notice that, when the temperature is lowered from 
$T=\infty$ down to $T=0$, the magnetization increases smoothly from
$m=0$ up to $m=1$. There is no abrupt change (singularity or non 
analyticity) in $m$ as a 
function of $\beta$ and therefore no phase transition.
\vskip .5cm
\noindent
{\em 
{\bf Exercise 1:} Consider two spins $\sigma_1$ and $\sigma_2$ 
with energy function 
\begin{equation}
E(\sigma_1,\sigma_2) = - \sigma_1 \sigma_2 \quad .
\end{equation}
Calculate the partition function, the magnetization of each spin as well as
the average value of the energy. Repeat these calculations for  
\begin{equation}
E(\sigma_1,\sigma_2) = - \sigma_1 - \sigma_2 \quad .
\end{equation}
How is the latter choice related to the single spin case?}

\subsubsection{Combinatorial meaning of the partition function.}

We have so far introduced statistical mechanics in probabilistic terms. 
There exists also a close relationship with combinatorics through the 
enumeration of configurations at a given energy; we now
show this relationship.

The average value of the energy may be computed directly from the definition
\begin{equation}
\langle E \rangle _T = \sum _C p (C) \; E(C) \quad ,
\end{equation}
or from the partition function $Z$ via the following identity
\begin{equation}
\label{ret5}
\langle E \rangle _T= - \frac{d}{d \beta } \ln Z\quad , \label{emoy}
\end{equation}
that can easily derived from (\ref{foncpart}). The identity (\ref{emoy}) 
can be extended to higher moments of the energy. For instance, the 
variance of $E$ can be computed from the second derivative of the 
partition function
\begin{equation}
\langle E ^2 \rangle _T - \langle E \rangle _T ^2 =  \frac{d^2}{d\beta^2} 
\ln Z \quad . \label{esqu}
\end{equation}
Such equalities suggest that $Z$ is the {\em generating function} of the 
configuration energies. To prove this statement, let us rewrite 
(\ref{foncpart}) as
\begin{eqnarray}
Z &=& \sum _C \exp \left( -\beta \; E({ C } )
\right) \nonumber \\
&=& \sum _{E}  { N} (E) \;\exp \left( -\beta \; E \right) 
\qquad ,
\label{foncpart2}
\end{eqnarray}
where ${ N}(E)$ is the number of configurations $C$ having
energies $E(C)$ precisely equal to $E$. If $x=\exp(-\beta)$, 
$Z(x)$ is simply the generating function of the coefficients 
${ N}(E)$ as usually defined in combinatorics.

The quantity $\hat S(E)=\ln { N}(E)$ is called the {\em entropy} 
associated with
the energy $E$. In general, calculating $\hat S(E)$ is a very hard
task. Usually, it is much more convenient to define the average entropy 
$\langle S \rangle _T$ at 
temperature $T$ as the contribution to the partition function which is
not directly due to energy,
\begin{equation}
\langle S \rangle _T = -\frac 1T \; 
\bigg( F(T) - \langle E \rangle _T \bigg) \quad ,
\label{entro}
\end{equation}
where 
\begin{equation}
F(T)=-T \ln Z(T) 
\label{deffree}
\end{equation}
is called the {\em free-energy} of the system.

In general, the above definitions for the energy and temperature
dependent entropies do not coincide. However, as explained in next
Section, in the large size limit $\langle S
\rangle _T$ equals $\hat S(E)$ provided that the energy $E$ is set to its
thermal average $E= \langle E \rangle _T$. 

The entropy is an increasing function of temperature. At
zero temperature, it corresponds to the logarithm of the number of absolute
minima of the energy function $E(C)$.
\vskip .5cm
\noindent
{\em {\bf Exercise 2:} Prove this last statement.}

\subsubsection{Large size limit and onset of singularity.}

We have not encountered any phase transition in the above examples
of systems with one or two spins. A necessary condition for the existence 
of a transition in a system is indeed that the size of the latter goes to 
infinity. The mathematical reason is simple: if the number of terms in the
sum (\ref{foncpart}) is finite, the partition function $Z$, the free-energy 
$F$, the average energy, ... are analytic functions of the inverse 
temperature $\beta$ and so do not have singularities at finite temperature. 

Most analytical studies are therefore devoted to the understanding of
the emergence of singularities in the free-energy when the size of the
system goes to infinity, the so-called thermodynamic limit. 

An important feature of the thermodynamic limit is the concentration
of measure for observables e.g. energy or entropy. Such quantities do
not fluctuate much around their mean values. More precisely, if we
call $N$ the size, {\em i.e.} the number of spins, of the system,
the moments of the energy usually scale as
\begin{eqnarray}
\langle E \rangle _T &=& O(N) \nonumber \\
\langle E ^2 \rangle _T - \langle E \rangle _T ^2 &=& O(N)
\qquad ,
\end{eqnarray}
and, thus the energy of a configuration is with high probability equal
to the average value up to $O(\sqrt N)$ fluctuations. Such a result
also applies to the entropy, and $\langle S \rangle _T = \hat S (
\langle E \rangle _T )$ up to $O(\sqrt N )$ terms. Measure
concentration in the thermodynamic limit is a very important and 
useful property, see \cite{Steele97}.

\subsection{Spin model on the complete graph.}\label{dep}

We shall now study a system of $N$ spins, called the Ising model, exhibiting 
a phase transition in the limit $N \to \infty$. We consider the complete
graph $K_N$; each vertex is labelled by an integer number 
$i=1,\ldots , N$ and carries a binary spin $\sigma
_i$. The energy function of a configuration 
$C=\{\sigma _1 , \ldots , \sigma _N \}$ is given by
\begin{equation}
E ( \sigma _1 , \ldots , \sigma _N ) = - \frac 1N \sum _{i<j}
\sigma _i \sigma _j - h \sum _i \sigma _i \qquad .
\label{chm1}
\end{equation}

\subsubsection{Remarks on the energy function.}

The first term in (\ref{chm1}) is called the interaction term. 
The sum runs over all
pairs of spins, that is over all edges of $K_N$. The minus sign ensures that
the minimum of energy is reached when all spins point in the same direction.
This direction depends on the second term of (\ref{chm1}) and, more precisely,
upon the sign of the ``magnetic field'' $h$. If the latter is positive
(respectively negative), the ground state is obtained when all spins are up 
(resp. down).

In the absence of field ($h=0$), we know the two ground states. The energy and 
entropy at zero temperature can be computed from (\ref{chm1}) and (\ref{entro}),
\begin{eqnarray}
\langle E \rangle _{T=0} &=& -\frac 12 (N-1)\quad ,\\
\label{chm2}
\langle S \rangle _{T=0} &=& \ln 2 \qquad .
\label{chm3}
\end{eqnarray}
Notice that the ground state energy is $O(N)$ due to the presence of the
factor $1/N$ in (\ref{chm1}) whereas the entropy is $O(1)$.

At infinite temperature, all configurations are equiprobable. The partition
function is simply equal to the total number of configurations:
$Z_{T=\infty} = 2^N$, leading to
\begin{eqnarray}
\langle E \rangle _{T=\infty } &=& 0 \quad , \\
\label{chm4}
\langle S \rangle _{T=\infty } &=& N\; \ln 2 \qquad .
\label{chm5}
\end{eqnarray}

When the temperature is finite, a compromise is realized in (\ref{foncpart2})
between energy and entropy: the {\it configurations} with low energies 
$E$ have the largest probabilities but the most probable energy also
depends on the entropy, {\em i.e.} on the size of the
coefficients ${ N}(E)$. Temperature tunes the relative 
importance of these two opposite effects. The phase transition studied in this 
section separates two regimes:
\begin{itemize}
\item a high temperature phase where entropy effects are dominant:
spins configurations are disordered and spins do not point in any
priviledged direction (for $h=0$). The average magnetization $m$ vanishes.
\item a low temperature phase where energy effects dominate: spins have 
a tendency to align with each other, resulting in ordered configurations 
with a non zero magnetization $m = \langle \sigma _i \rangle _T \ne 0$.
\end{itemize}
Let us stress that the energy and the entropy must have the same orders of
magnitude (=$O(N)$) to allow for such a compromise and thus for 
the existence of  a phase transition at finite strictly positive
temperature.

\subsubsection{The magnetization is the order parameter.} 

\begin{figure}
\begin{center}
\resizebox{0.35\linewidth}{!}{\includegraphics{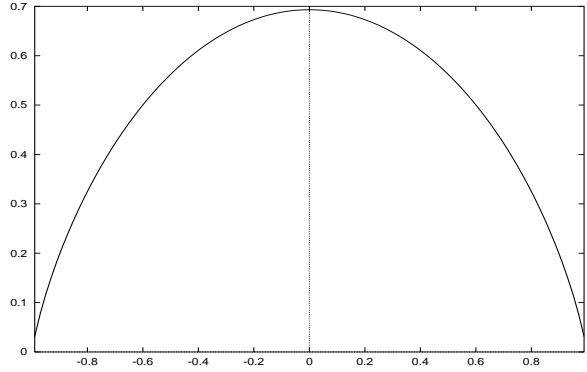}}
\end{center}
\caption{Entropy $s(m)$ of the Ising model on the complete graph
as a function of magnetization $m$.}
\protect\label{isingentro}
\end{figure}

We start by defining the magnetization of a configuration $C
=\{\sigma _1 , \ldots , \sigma _N \}$ as
\begin{equation}
m ( C ) = \frac 1N \sum _{i=1} ^N \sigma _i
\quad .
\label{aimconf}
\end{equation}
The calculation of the partition function relies on the following remark.
The energy function (\ref{chm1}) depends on the configuration $C$ 
through its magnetization $m ( C ) $ only. More 
precisely, 
\begin{equation}
E ( C ) =  - N \left( \frac 12 \; m ( C ) ^2 + h 
\; m ( C ) \right) + \frac 12
\quad .
\label{aim2}
\end{equation}
In the following, we shall also need
the entropy at fixed magnetization $S(m)$.
Configurations with a fixed magnetization $m$ have $N_+$ spins up and
$N_-$ spins down with
\begin{eqnarray}
N_+ &=& N \; \left( \frac {1+m}2 \right) \quad , \nonumber \\
N_- &=& N \; \left( \frac {1-m}2 \right) \quad .
\end{eqnarray}
The number of such configurations is therefore given by the binomial 
coefficient 
\begin{equation}
e^{S(m)} = \frac{N!}{N_+ ! N_- !} \qquad .
\end{equation}
In the large $N$ limit, Stirling's formula gives access to the asymptotic
expression of the entropy density, $s(m)=S(m)/N$, at fixed magnetization,
\begin{equation}
s(m) = - \left( \frac {1-m}2 \right) \; \ln \left( \frac {1-m}2 \right) 
- \left( \frac {1+m}2 \right) \; \ln \left( \frac {1+m}2 \right)   
\qquad , \label{hg}
\end{equation}
Figure~\ref{isingentro} displays $s(m)$ as a function of $m$. The maximum
is reached at zero magnetization ($s(0)=\ln 2$) and the entropy vanishes on
the boundaries $m=\pm 1$. 

Let us stress that $S(m)$ defined in (\ref{hg}) is the entropy at given
magnetization and differs {\em a priori} 
from the energy and temperature dependent
entropies, $\hat S (E)$ and $\langle S \rangle _T$, defined above.
However, in the thermodynamic limit, all quantities are equal provided
that $m$ and $E$ coincide with their thermal averages, $\langle m
\rangle _T$ and $\langle E \rangle _T$.

The average value $\langle m \rangle _T$ of the magnetization will be
shown to vanish in the high temperature phase and to be different from
zero in the low temperature phase. The magnetization is an order
parameter: its value (zero or non-zero) indicates in which phase the
system is.

\subsubsection{Calculation of the free-energy.}

The partition function $Z$ reads
\begin{eqnarray}
Z&=& \sum _{ \sigma _1 , \ldots , \sigma _N =\pm 1 }
\exp \left[ - \beta \; E ( \sigma _1 , \ldots , \sigma _N ) \right]
\nonumber \\
&=& \sum _{ m =-1,-1+\frac 2N , \ldots , 1 - \frac 2N , 1}
\exp \left[ - N \; \beta \; \hat f(m) \right] \qquad ,
\label{foncpart3}
\end{eqnarray}
where 
\begin{equation}
\hat f (m) =  -\frac 12 \; m ^2 - h 
\; m  - T \; s(m) \qquad , \label{isingf}
\end{equation}
up to  $O(1/N)$ terms. For the moment, we shall take $h=0$.

\begin{figure}
\begin{center}
\resizebox{0.3\linewidth}{!}{\includegraphics{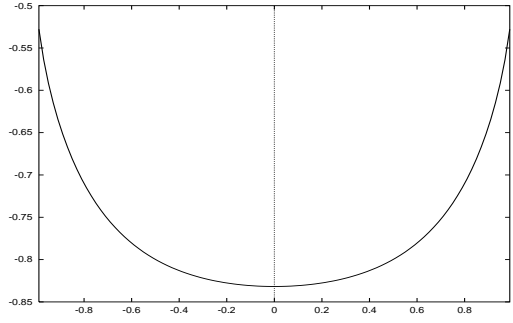}}
\\
\resizebox{0.3\linewidth}{!}{\includegraphics{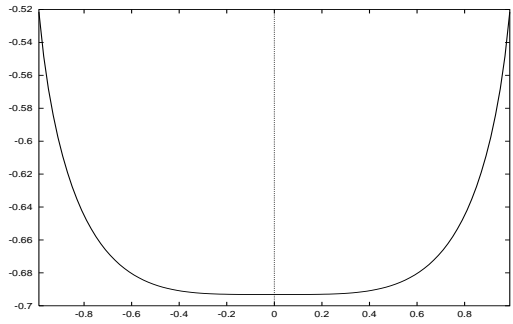}}

\resizebox{0.3\linewidth}{!}{\includegraphics{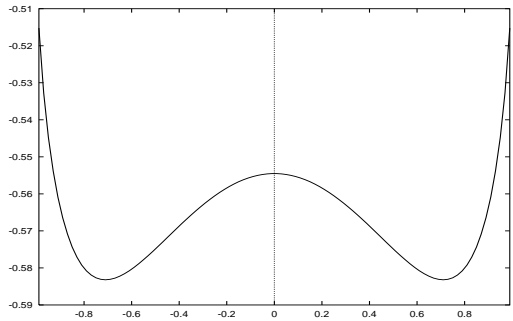}}
\end{center}
\caption{Free-energy function $\hat f(m)$ of the Ising model on the complete
graph  as a function of the magnetization $m$ in zero magnetic field $h$
and for three different temperatures. {\bf a}:~high
temperature $T=1.2$, {\bf b}:~critical temperature $T=1$, 
{\bf c}:~low temperature $T=0.8$.} 
\protect\label{isingelib}
\end{figure}

In the limit of an infinite number $N$ of spins, the free-energy may
be computed by means of the saddle-point (Laplace) method. We look for 
the saddle-point magnetization $m^*$ (that depends upon temperature $T$)
minimizing $\hat f (m)$ (\ref{isingf}). The latter is plotted 
in Figure~\ref{isingelib} for three different temperatures. 
 
It can be seen graphically that the minimum of $\hat f$ is located at
$m^*=0$ when the temperature is larger than $T_c=1$ 
while there exist two opposite minima, $m=-m^*(T) < 0$, $m=m^* (T)> 0$ 
below this critical temperature. The optimum magnetization is solution of
the saddle-point equation,
\begin{equation}
m^* = \tanh \big( \beta \; m^* \big) \qquad ,
\label{isingcol}
\end{equation}
while the free-energy is given by
\begin{equation}
f(T)= \lim _{N\to\infty} - \frac TN \ln Z = \hat f( m^*)  \qquad . 
\label{ret4}
\end{equation}
The average energy and entropy per spin (divided by $N$) can be computed from 
(\ref{ret4}, \ref{ret5}, \ref{entro}), 
\begin{eqnarray}
\langle e \rangle _{T } &=& -\frac 12 \; (m^* )^2 \quad , \\
\label{isener}
\langle s \rangle _{T } &=& s ( m^* ) \qquad .
\label{isentr}
\end{eqnarray}

\subsubsection{Phase transition and symmetry breaking.}

In the absence of a magnetic field, the energy 
(\ref{chm1}) is an even function of the spins: the probability of
two opposite configurations $\{\sigma _1 , \ldots , \sigma _N \}$  and
$\{-\sigma _1 , \ldots , - \sigma _N \}$ are equal. As a consequence,
the thermal average $\langle \sigma \rangle _T$  of any spin vanishes.
This result is true for any $N$ and so, in the large $N$ limit,
\begin{equation}
\lim _{N\to\infty} \; \lim _{h \to 0} \; \langle \sigma
\rangle _T = 0 \quad .
\label{ok}
\end{equation}
It is thus necessary to unveil the meaning of the saddle-point
magnetization $m^*$ arising in the computation of the partition function.

To do so, we repeat the previous calculation of the free-energy in 
presence of a magnetic field $h>0$. The magnetization is now different from
zero. At high temperature $T> T_c$, this magnetization decreases
as the magnetic field $h$ is lowered and vanishes when $h=0$,
\begin{equation}
\lim _{h \to 0^+} \;  \lim _{N\to\infty}  \; \langle \sigma
\rangle _T = 0 \qquad (T>T_c )\quad .
\label{ok1}
\end{equation}
Therefore, at high temperature, the inversion of limits between
(\ref{ok}) and (\ref{ok1}) has no effect on the final result.

The situation drastically changes at low temperature. When $T <T_c$,
the degeneracy between the two minima of $f$ is lifted by the
magnetic field. Due to the field, a contribution $-h\; m$ must be added
to the free-energy (\ref{isingf}) and favours the minimum in $m^*$ over
that in $-m^*$. The contribution to the partition function 
(\ref{foncpart3}) coming from the second minimum is exponentially
smaller than the contribution due to the global minimum in $m^*$
by a factor $\exp ( - 2 N \beta h m^* )$. The probability measure on 
spins configurations is therefore fully concentrated around the global
minimum with positive magnetization and
\begin{equation}
\lim _{h \to 0^+} \;  \lim _{N\to\infty}  \; \langle \sigma
\rangle _T = m^* \qquad (T<T_c )\quad .
\label{ok2}
\end{equation}
From (\ref{ok}) and (\ref{ok2}), the meaning of the phase transition 
is now clear. Above the critical temperature, a small perturbation
of the system (e.g. a term in the energy function pushing spins 
up), is irrelevant: as the perturbation disappears ($h\to 0$), 
so do its effects ($m^*\to 0$), see~(\ref{ok1}). Conversely, below the
critical temperature, a small perturbation is enough to trigger 
strong effects: spins point up (with a
spontaneous magnetization $m^* > 0$) even after the
perturbation has disappeared ($h=0$), see~(\ref{ok2}). At low temperature,
two phases with opposite magnetizations $m^*$ and $-m^*$ coexist.
Adding an infinitesimal field $h$ favours and selects 
one of them. In more mathematical terms, the magnetization $m$ is a
non-analytic and discontinuous function of $h$ at $h=0$. 

So, the phase transition here appears 
to be intimately related to the notion of {\em symmetry breaking}. 
In the case of the Ising model, the probability distribution
over configurations is symmetrical, that is, left unchanged 
under the reversal of spins $\sigma \to - \sigma$. A high temperature, 
this symmetry also holds for average quantities: 
$\langle \sigma \rangle _T =0$. At low temperature, the reversal
symmetry is broken since, in presence of an infinitesimal
perturbation, $\langle 
\sigma \rangle _T = m^* \ne 0$. The initial symmetry of the system
implies only that the two possible phases of the system have opposite
magnetizations $m^* $ and $-m^*$.

In the present case, the symmetry of the system was easy to identify,
and to break! We shall see that more abstract and complex symmetries may
arise in other problems, e.g. the random graph and K-Satisfiability.
The understanding of phase transitions very often will rely 
on the breaking of associated symmetries.

\vskip .5cm
\noindent
{\em 
{\bf Exercise 3:} How does equation (\ref{isingcol}) become modified 
when there is a non-zero magnetic field?
Calculate explicitely the free-energy in presence of a
magnetic field and check the correctness of the above statements.
}

\subsubsection{Vicinity of the transition and critical exponents.}
\label{critexpo1}

To complete the present analysis, we now investigate the properties 
of the Ising model close to the critical temperature $T_c=1$ and define
$T=1+\tau$ with $|\tau|\ll 1$. The spontaneous magnetization reads from
(\ref{isingcol}),
\begin{equation}
m^* (\tau ) = \cases{ 0 & if \hbox{$\tau \ge 0$} \quad , \cr
\sqrt{- 3\tau } & if \hbox{$\tau \le 0$} \quad . \cr }
\end{equation}
Thus the magnetization grows as a power of the shifted temperature $\tau$: 
$ m^* (\tau ) \sim (-\tau )^\beta$ with $\beta = 1/2$. $\beta$, not
to be confused with the inverse temperature, is called a {\em critical 
exponent} since it characterizes the power law behaviour of a physical
quantity, here the magnetization, close to criticality. Such exponents
are universal in that they are largely independent of the ``details'' of 
the definition of the model. We shall come back to this point in the 
sections devoted to the random graph and the K-Satisfiability models.

Another exponent of interest is related to the finite size effect
at the transition. So far, we have calculated the average values of
various quantities in the infinite size limit $N\to\infty$. We have
in particular shown the existence of a critical temperature 
separating a phase where the sum of the spins is on average zero
$(\tau >0$) from a phase where the sum of the spins acquires an 
$O(N)$ mean  ($\tau < 0$). At the transition point
($\tau =0$), we know that the
sum of spins cannot be of order $N$; instead we have a
scaling in $N^\delta$ with $\delta <1$. 

What is the value of  $\delta$? From expression (\ref{foncpart3}), let us
expand the free-energy function $\hat f(m)$ (\ref{isingf}) in powers of the
magnetization $m=O(N^{\delta  -1 })$,
\begin{equation} \label{deve}
f(m) - f(0) = \frac {\tau}2 \; m ^2 +\frac 1{12}\; m^4  +
O( m^6 , \tau \; m^4 ) \qquad , \label{isingf2}
\end{equation}
with $f(0)=-T\ln 2$. Above the critical temperature, $\tau >0$, the
average magnetization is expected to vanish. Due to the presence of the
quadratic leading term in (\ref{deve}), the fluctuations of $m$ are
of the order of $N^{-1/2}$. The sum of the spins, $N\, m$, has a 
distribution whose width grows as $N^{1/2}$, giving  $\delta=1/2$.

At the critical temperature, the partition function
reads from (\ref{foncpart3}),
\begin{equation}
Z \simeq 2^N \; \int dm \; e^{- N\; m^4 / 12 } \qquad . 
\label{foncpart4}
\end{equation}
The average magnetization thus vanishes as expected and fluctuations are 
of the order of $N^{-1/4}$. The sum of the spins, $N\, m$, thus
has a distribution whose width grows as
$N^{3/4}$, giving  $\delta=3/4$.

The {\em size} of the critical region (in temperature) is defined 
as the largest value $\tau _{max}$ of the shifted temperature 
$\tau$ leaving unchanged the order of magnitude of the fluctuations of 
the magnetization $m$. A new critical exponent $\nu$ that monitors this
shift is introduced: $\tau _{max} \sim N^{-1/\nu}$. Demanding
that terms on the r.h.s. of (\ref{isingf2}) be of the same order in $N$, 
we find $\nu =2$.

\subsection{Randomness and the replica method.}

The above analysis of the Ising model has been useful to illustrate some
classic analytical techniques and to clarify the concept of phase 
transitions. However, most optimization or decision problems encountered
in computer science contain another essential ingredient we have not 
discussed so far, namely randomness. To avoid any confusion, let us stress
that randomness in this case, e.g. a Boolean {\em formula} randomly drawn from
a well-defined distribution, and called {\em quenched disorder} in
physics, must be clearly distinguished from the 
probabilistic formulation of statistical mechanics related to the
existence of {\em thermal disorder}, see (\ref{gibbs}). As already
stressed, as far as combinatorial aspects of statistical mechanics are
concerned, we can start from the definition (\ref{foncpart2})
of the partition function and interpret it as a generating function, 
forgetting the probabilistic origin. On the contrary, quenched disorder 
cannot be omitted.
We are then left with combinatorial problems defined on
random structures, that is, with partition functions where the weights
themselves are random variables.

\subsubsection{Distribution of ``quenched'' disorder.}

We start with a simple case:
\vskip .3cm
\noindent
{\em 
{\bf Exercise 4:} Consider two spins $\sigma_1$ and $\sigma_2$ 
with energy function 
\begin{equation}
E(\sigma_1,\sigma_2) = - J \; \sigma_1 \; \sigma_2 \quad ,
\end{equation}
where $J$ is a real variable called coupling.
Calculate the partition function, the magnetization of each spin as well as
the average value of the energy at given (quenched) $J$. 
Assume now that the coupling $J$ is
a random variable with measure $\rho (J)$ on a finite support $[J_-;
J_+]$. Write down the expressions of the mean over $J$ of the 
magnetization and energy. What is the value of the average ground state
energy? }

The meaning of the word ``quenched'' is clear from the above example. 
Spins are always distributed according to (\ref{gibbs}) but the energy
function $E$ now depends on randomly drawn variables e.g. the coupling
$J$. Average quantities (over the probability distribution
$p$) must be computed keeping these
random variables fixed (or quenched) and thus are random variables
themselves that will be averaged over $J$ later on. To distinguish both
kinds of averages we hereafter
use an overbar to denote the average over the quenched random
variables while brackets still indicate a thermal average using $p$.

Models with quenched randomness are often very difficult to solve.
One of the reasons is that their physical behaviour is more complex 
due to the presence of {\em frustration}.

\subsubsection{Notion of frustration.}
\label{refsk}

Frustration is best introduced through the following simple example.
\vskip .3cm
{\em 
{\bf Exercise 5:} Consider three spins $\sigma_1$, $\sigma_2$ 
and $\sigma _3$ with energy function 
\begin{equation}
E(\sigma_1,\sigma_2 , \sigma _3) = - \sigma_1 \sigma_2 - \sigma_1 \sigma_3 -
 \sigma_2 \sigma_3 \quad .
\end{equation}
Calculate the partition function, the magnetization of each spin as well as
the average value of the energy. What are the ground state energy and entropy?

Repeat the calculation and answer the same questions for
\begin{equation} \label{inut}
E(\sigma_1,\sigma_2, \sigma _3) = - \sigma_1 \sigma_2 - \sigma_1 \sigma_3 +
 \sigma_2 \sigma_3 \quad .
\end{equation}
Note the change of the last sign on the r.h.s. of (\ref{inut}).}

The presence of quenched disorder with both negative and positive couplings
generates frustration, that is conflicting terms in
the energy function. A famous example is the Sherrington-Kirkpatrick (SK)
model, a random version of the Ising model on the complete graph whose
energy function reads
\begin{equation}
E _{SK} (\sigma _1 ,\ldots , \sigma _N ) = - \frac 1{\sqrt N} \sum _{i<j}
J_{ij} \sigma _i \sigma _j \qquad ,
\label{chmsk}
\end{equation}
where the quenched couplings $J_{ij}$ are independent random normal
variables. In the SK model, contrarily to the Ising model, the product
of the couplings $J_{ij}$ along the loops of the complete graph $K_N$
may be negative. The ground state is no longer given by the ``all spins
up'' configuration, nor by any simple prescription and must
be sought for among the set of $2^N$ possible configurations. Finding
the ground state energy for an arbitrary set of couplings $J_{ij}$
is a hard combinatorial optimization task which in this case
belongs to the class
of NP-hard problems~\cite{GareyJohnson79,PapadimitriouSteiglitz}. 

\subsubsection{Thermodynamic limit and self-averaging quantities.}

Though physical quantities depend {\em a priori} on quenched couplings,
some simplifications may take place in the large size limit 
$N\to\infty$. Many quantities of interest 
may exhibit less and less fluctuations around their mean values and
become {\em self-averaging}. In other words, the distributions of some
random variables become highly concentrated as $N$ grows. Typical examples
of highly concentrated quantities are the (free-)energy, the entropy, the
magnetization, ... whereas the partition function is generally not
self-averaging.

Self-averaging
properties are particularly relevant when analyzing a problem. Indeed,
for these quantities, 
we only have to compute their average values,
not their full probability distributions. We shall encounter numerous
examples of concentrated random variables later in this article.
\vskip .3cm
{\em 
{\bf Exercise 6:}
Show that the partition function of the SK model is not self-averaging
 by calculating its first two moments.}

\subsubsection{Replica method.}

We consider a generic model with $N$ spins $\sigma _i$ and an energy function
$E( C , J )$ depending on a set of random couplings $J$.
Furthermore we assume that the free-energy $F(J)$ of this model is 
self-averaging and would like to compute its quenched averaged value 
$\overline{F(J)}$ or, equivalently from (\ref{deffree}), the averaged 
logarithm of the partition function $\overline{\ln Z(J)}$. 
Though well posed, this computation is generally
a very hard task from the analytical point of view. An original but
non rigorous method, the {\em replica approach}, was invented by Kac 
in the sixties to perform such calculations. 
The starting point of the replica approach is the following expansion
\begin{equation}
Z(J) ^n = 1 + n \; \ln Z(J) + O(n^2)
\qquad ,
\label{replimeth}
\end{equation}
valid for any set of couplings $J$ and small real $n$. 
The identity (\ref{replimeth}) may be averaged over couplings and gives
the mean free-energy from the averaged n$^{th}$ power of the partition
function
\begin{equation}
\overline{F(J)} = -T \; \lim _{n\to 0} \left( \frac{  \overline{Z(J) ^n }
-1 } {n} \right) \quad .
\label{replimeth2}
\end{equation}
If we restrict to {\em integer} $n$, the n$^{th}$ moment of the 
partition function $Z$ can be rewritten as 
\begin{eqnarray}
\overline{Z(J) ^n } &=& \overline{ \left[ 
\sum _C \exp \left( -\frac 1T \; E({ C }, J ) \right) \right] ^n }
\nonumber \\
&=&  \sum _{C^1 , \ldots ,  C^n } 
\overline{  \exp \left( -\frac 1T \; \sum _{a=1}
^n E( { C }^a , J ) \right)  } \quad .
\label{repl56}
\end{eqnarray}
This last expression makes transparent the principle of the replica method.
We have $n$ copies, or replicas, of the initial problem.
The random couplings disappear once the average over the quenched couplings
has been carried out. Finally, we must compute the partition function
of an abstract system of $N$ vectorial spins $\vec \sigma _i = (
\sigma _i ^1 , \ldots , \sigma _i ^n )$ with the non random 
energy function
\begin{equation}
E_{eff} (  \{\vec \sigma _i \} ) = - T \; \ln \left[ 
\overline{  \exp \left( -\frac 1T \; \sum _{a=1}
^n E( { C }^a , J ) \right)  } \right] \quad .
\label{repl57}
\end{equation}
This new partition function can be estimated 
analytically in some cases by means
of the saddle-point method just as we did for the Ising model. 
The result may be written formally as
\begin{equation}
\overline{Z(J) ^n } = \exp \bigg( - N  \tilde f (n) \bigg)\quad ,
\label{replimeth3}
\end{equation}
to leading order in $N$. On general grounds, there is no reason to
expect the partition function to be highly concentrated. Thus, $\tilde
f (n)$ is a non linear function of its integer argument $n$ satisfying
$\tilde f (0)=0$.  The core idea of the replica approach is to continue
analytically $\tilde f$ to the set of real $n$ and obtain
$\overline{F(J)}=T N d{\tilde f}/dn$ evaluated at $n=0$. 
The existence and uniqueness of
the analytic continuation is generally ensured for finite sizes $N$
due to the moment theorem. In most problems indeed one succeeds in
bounding $|Z(J)|$ from above by a (J independent) constant $C$. The
moments of $Z$ grow only exponentially with $n$ and their knowledge
allows for a complete reconstruction of the probability distribution
of $Z(J)$.  However this argument breaks down when the saddle-point
method is employed and the upper bound $C = \exp (O(N))$ becomes
infinite.

Though there is generally no rigorous scheme for the analytic
continuation when $N\to \infty$, physicists
have developped in the past twenty years many empirical rules to use
the replica method and obtain precise and sometimes exact results for
the averaged free-energy. We shall see in the case of the K-Satisfiability
problem how the replica approach can be applied and how very peculiar
phase transitions, related to the abstract ``replica'' 
symmetry breaking, are present.

The mathematician or computer scientist reader of this
brief presentation may feel uneasy and distrustful of the 
replica method because of the 
uncontrolled analytic continuation. To help him/her 
loose some inhibitions, he/she is asked to consider the following
warming up exercise:
\vskip .3cm
{\em 
{\bf Exercise 7:}
Consider Newton's binomial expression for $(1+x)^n$ with
integer $n$ and perform an analytic continuation to real $n$.
Take the $n \to 0$ limit and show that this leads to the
series expansion in $x$ of $\ln (1+x)$. }

\section{Random Graphs}
\label{randgraph}

In this section, we show how the statistical mechanics concepts and
techniques exposed in the previous section allow to reproduce some
famous results of Erd\"os and R\'enyi on random graphs\cite{Bollo}.

\subsection{Generalities}

First let us define the random graphs used.
Consider the complete graph $K_N$ over $N$ vertices.
We define ${ G}_ {N,N_{L}}$ as the set of graphs obtained by taking only
$N_{L}=\gamma \; N / 2$  among the ${N\choose 2}$ edges of $K_N$ in all 
possible different ways.
A {\sl random graph} is a randomly chosen element of 
${ G}_{N,N_{L}}$ with the flat measure.
Other random graphs can be generated from the complete graph $K_N$ 
through a random deletion process of the edges
with probability $1- \gamma/N$. In the large $N$ limit, both
families of random graphs share common properties and we shall mention
explicitely the precise family we use only when necessary.

\subsubsection{Connected components.}

We call ``clusters'' the connected components of a given
graph $G$; the ``size'' of a cluster is
the number of vertices it contains.  An
isolated vertex is a cluster of size unity.  The number of connected
components of $G$ is denoted by $C(G)$ and we shall indicate its normalized
fraction by $c(G) =\frac C{N}$.  If $c$ is small, the random graph
$G$ has few big clusters whereas for $c$ approaching unity
there are many clusters of
small size.  Percolation theory is concerned with the study of the
relationship between the probability $p$ of two vertices being
connected with the typical value of $c $ in the $N \to \infty$ limit.
The scope of this section is to show how such a relationship can be
exploited by the study of a statistical mechanics model, the
so called Potts model, after a suitable analytic
continuation. As a historical note, let us mention that analytic
continuations have played an enormous role in  physics this last
century, leading often to unexpected deep results,
impossible or very difficult to obtain by other means.

\subsubsection{Generating function for clusters.}

Let ${\mathcal{P}}(G)$ be the probability of
drawing a random graph $ G$ through the deletion process
from the complete graph $K_N$.  Since the edge deletions are
statistically independent, this probability depends on the number of
edges $N_L$ only, and factorizes as
\begin{equation}
{\mathcal{P}}(G)=
\mathit{p} ^{N_{\mathit{L}}(G)}
\mathit{(1-p)}^{\frac{N(N-1)}{2}-N_{\mathit{L}}(G)}
\qquad ,
\end{equation}
where
\begin{equation}
1-p = 1-\frac{ \gamma}{N}
\label{prodel}
\end{equation}
is the probability of edge deletion.
We want to study the probability density $\rho (c)$ of generating a
random graph with $c$ clusters,
\begin{equation}
\rho (c)=\sum_{G}{\mathcal{P}}(G) \;
\delta(\mathit{c}-\mathit{c}(G)) \qquad ,
\label{probn}
\end{equation}
where $\delta$ indicates the Dirac distribution. 

We can introduce a generating function of the cluster probability by
\begin{eqnarray}
Y (q)  &=& \int_{0}^{1} dc\,\rho (c ) q^{N\; c}
\nonumber \\ &=& \int _0 ^1 dc\, q^{N\;c}\sum_{G \subseteq K_N }
{\mathcal{P}}(G)\delta(\mathit{c}-\mathit{c} (G) )\nonumber \\
&=&\sum_{G\subseteq K_N}{\mathcal{P}}
(G)\mathit{q}^{ C(G)}=
\sum_{G\subseteq K_N} p^{L
(G)}(1-p)^{\frac{N(N-1)}{2}-L(G)} q^{C(G)} \ ,
\label{generatricesum}
\end{eqnarray}
with $q$ being a formal (eventually real) parameter.

\subsubsection{Large size limit.}

In the large size limit, $\rho (c)$ is expected to be highly
concentrated around some value $c(\gamma )$ equal to
 the typical fraction of clusters per vertex and depending only the
average degree of valency $\gamma$. Random graphs whose $c(G)$ differs enough
from $c(\gamma )$ will be exponentially rare in
$N$. Therefore, the quantity
\begin{equation}
\omega(c)=\lim _{N\to \infty} \; \frac{1}{N}\log {\rho}(c) 
\label{omesad}
\end{equation}
should vanish for $c=c(\gamma )$ and be strictly negative 
otherwise. In the following, we shall compute $\omega (c)$ and thus
obtain information not only on the typical number of clusters but also
on the large deviations (rare events).  

Defining the logarithm $\tilde f (q)$ of the cluster generating function
as
\begin{equation}
\tilde f (q) =\lim _{N\to \infty} \; \frac{1}{N}\log Y(q) \qquad , 
\end{equation}
we obtain from a saddle-point calculation on $c$, see 
(\ref{generatricesum},\ref{omesad}),
\begin{equation}
\tilde f (q)=\max _{0\le c\le 1} \bigg[ c \;\ln q + \omega (c) \bigg] \qquad .
\label{generatriceint}
\end{equation}
In other words, $\tilde f$ and $\omega$ are simply conjugated
Legendre transforms. It turns out that a direct computation of
$\tilde f$ is easier and thus prefered.

\subsection{Statistical mechanics of the random graph.}
 
Hereafter, we proceed to compute the properties of random graphs by
using a mapping to the so-called Potts model. Some know results can be 
rederived by the statistical mechanics approach, and additional
predictions are made.

\subsubsection{Presentation of the Potts model.}

The Potts model\cite{Potts52} is defined in terms of an energy function
which depends on $N$ spin variables $\sigma_{i}$, one for each vertex
of the complete graph $K_N$, which take $q$ distinct 
values  $\sigma_{i}=0,1,...,q-1$.
The energy function reads
\begin{equation}
E[ \{ \sigma _i \} ]=- \sum_{ i<j }\delta(\sigma_{i},\sigma_{j}) \; \; ,
\label{hampotts}
\end{equation}
where $\delta(a,b)$ is the Kronecker delta function:
$ \delta(a,b)=1$ if $a=b$ and   $ \delta(a,b)=0$ if $a \neq b$.
The partition function of the Potts model is
\begin{equation}
Z_{Potts}= \sum_{\{\sigma_{i}=0,...,q-1\}}
\exp[\beta \sum_{ i< j}\delta(\sigma_{i},\sigma_{j})]
\label{partizpotts}
\end{equation}
where $\beta$ is the inverse temperature
and the summation runs over all $q^N$ spin configurations.

In order to identify the mapping between the statistical mechanics features 
of the Potts model and the percolation problem in random graphs we 
compare the expansion of $Z_{Potts}$ to the definition of the 
cluster generating function of the random graphs.

\subsubsection{Expansion of the Potts partition function.}

Following  Kasteleyn and Fortuin \cite{Kasteleyn}, we start by rewriting 
$Z_{Potts}$ as a dichromatic polynomial. Upon posing
\begin{equation}
v=e^{\beta }-1 \; \;,
\end{equation}
one can easily check that (\ref{partizpotts}) can be recast in the form
\begin{equation}
Z_{Potts}=\sum_{\{\sigma_{i}\}}\prod_{i<j }[1+v\delta(\sigma_{i},
\sigma_{j})] \; \;.
\end{equation}
When $\sigma_i$ and $\sigma_j$ take the same value there appears a
factor $(1+v)$ in the product (corresponding to a term $e^{\beta}$ in
(\ref{partizpotts})); on the contrary, whenever $\sigma_i$ and $\sigma_j$
are different the product remains unaltered.
The expansion of the above product reads
\begin{eqnarray}
Z_{Potts}=\sum_{\{\sigma_{i}\}}[1&+&v\sum_{ i<j }
\delta(\sigma_{i},\sigma_{j}) \nonumber \\ &+& v^{2}
\sum_{ i<j ,  k< l /{(i,j)\ne (k,l)}}
\delta(\sigma_{i},\sigma_{j})\delta(\sigma_{k},\sigma_{l})+\cdots] \; .
\end{eqnarray}
We obtain $2^{\frac{N(N-1)}{2}}$ terms each of which composed by two factors,
the first one given by $v$ raised to a power equal to the number of $\delta$s
composing the second factor.
It follows that each term corresponds to a possible subset of edges
on $K_N$, each edge weighted by a factor $v$.
There is a one--to--one correspondence between each term of the sum
and the sub--graphs $G$ of $K_N$.
The edge structure of each sub--graph is encoded in the product of the 
$\delta$s.
This fact allows us to rewrite the partition function as a sum
over sub--graphs
\begin{equation}
Z_{Potts}=\sum_{\{\sigma_{i}\}}\sum_{G \subseteq K_N}
[v^{L(G )}
\prod_{k=0}^{L(G)}
\delta ( \sigma_{i_k},\sigma_{j_k} ) ]
\end{equation}
where $L(G)$ is the number of edges in the sub--graph $G$ and
${i_k},j_k$ are the vertices connected by the $k$-th edge of the
sub--graph.  We may now exchange the order of the summations and perform
the sum over the spin configurations.  Given a sub--graph $G$ with $L$
links and $C$ clusters (isolated vertices included), the sum over spins
configurations will give zero unless all the $\sigma$s belonging to a 
cluster of $G$ have the same value (cf. the $\delta$ functions).
In such a cluster, one can set the $\sigma$s to any of the $q$ different 
values and
hence the final form of the partition function reads

\begin{equation}
Z_{Potts}=\sum_{G\subseteq K_N}v^{L(G)}
q^{C(G)} \; \; .
\label{zpotts}
\end{equation}

\subsubsection{Connection with the cluster generating function}

If we now make the following identification
\begin{equation}
p=1-e^{-\beta }=v/(1+v)\; \;,
\end{equation}
we can rewrite the partition function as
\begin{eqnarray}
Z_{Potts}&=&\sum_{G\subseteq K_N} 
\biggl(\frac{p}{1-p}\biggr)^{L(G)}
q^{C(G)} \nonumber \\
&=&(1-p)^{-\frac{N(N-1)}{2}}\sum_{G\subseteq K_N}
p^{L(G)}(1-p)^{\frac{N(N-1)}{2}
-L(G)}q^{C(G)} \; \;.
\label{percola}
\end{eqnarray}
Computing the prefactor on the r.h.s. of (\ref{percola}), we have
\begin{equation}
Z_{Potts} =e^{\frac{N\gamma}{2}}\; Y(q) \; \;,
\end{equation}
for terms exponential in $N$. $Y$ is the cluster generating function of 
the graph  (\ref{generatricesum}). 
The large $N$ behaviour of the cluster probability $\omega (c)$ is
therefore related to the Potts free--energy,
\begin{equation}
f_{Potts} (q)=-\lim_{N \rightarrow \infty} \frac{1}{\beta \;N} 
\ln Z_{Potts} \; \;,
\label{pottsfe}
\end{equation}
through
\begin{equation}
-\frac{\gamma}{2}-f_{Potts}
(q)=\max _{0\le c \le 1} (c\; \ln q+\omega(c)) \; \;.
\label{colle}
\end{equation}
We are interested in finding the value $c^*(q)$ which maximizes
the r.h.s. in (\ref{colle}); since
\begin{equation}
\frac{d\omega (c)}{d c} \biggr|_{c^* (q)}=-\ln q
\label{logq=1}
\end{equation}
it follows that $\omega$ takes its maximum value for $q=1$.
Differentiating eq. (\ref{colle}) with respect to $q$, we have
\begin{equation}
-\frac{df_{Potts}}{dq}=\frac{d}{dq}(c \; \ln q + \omega(c))=
\frac{\partial}{\partial c}(c \; \ln q +
 \omega(c))\frac{\partial c}{\partial q}+\frac{c}{q} \; \;,
\end{equation}
which, in virtue of eq. (\ref{logq=1}) becomes:
\begin{equation}
c^*(q) = -q \; \frac{df _{Potts}}{dq} (q) \qquad .
\label{numclus}
\end{equation}

It is now clear that the typical fraction of clusters per site,
$c^*(q=1)$, can be obtained, at a given connectivity $\gamma$, by
computing the Potts free-energy in the vicinity of $q=1$. Since the
Potts model is originally defined for integer values of $q$ only, an
analytic continuation to real values of $q$ is necessary. We now explain
how to perform this continuation.

\subsubsection{Free-energy calculation.}

As in the case of the Ising model of section II, a careful
examination of the energy function (\ref{hampotts}) shows that the
latter depends on the spin configuration only through the fractions 
$x (\sigma ; \{\sigma _i \} )$ of variables $\sigma_i$ in the $\sigma$-th state
($\sigma=0,1,\cdots,q-1$)\cite{Wu82},
\begin{equation}
x({\sigma}; \{\sigma _i \})
=\frac{1}{N} \sum_{i=1}^N \delta(\sigma_i,\sigma), \quad
(\sigma=0,1,...,q-1) \qquad .
\end{equation}
Of course, $\sum_{\sigma} x({\sigma};\{\sigma _i \}
)=1$. Note that in the Ising case ($q=2$) the two fractions $x(0)$ 
and $x(1)$ can be parametrized by a unique parameter
e.g. the magnetization $m=(x(1)-x(0))/2$.

Using these fractions, the energy (\ref{hampotts}) may be rewritten as
\begin{equation}
E[\{\sigma_i\}]= -\frac{N^2}{2} \sum_{\sigma =0}^{q-1} [ x (\sigma ;
\{\sigma_i\} )]^2 + \frac N2 \qquad .
\label{Hx}
\end{equation}
Note that the last term on the r.h.s. of (\ref{Hx}) can
be neglected with respect to the first term whose order of magnitude is
$O(N^2)$.

The partition function (\ref{partizpotts}) 
at inverse temperature $\beta = \gamma /N$ now becomes
\begin{eqnarray}
Z_{Potts}  &=& \sum_{ \{ \sigma_i = 0,1,...q-1 \}} 
 \exp \left( -\frac{\gamma}2 \; N \sum_{\sigma=0}^{q-1} [
x(\sigma , \{\sigma_i\} )]^2 \right) \nonumber \\
 &=& \sum_{ \{ x_\sigma = 0,1/N,...,1 \}} ^{(R)}
 \exp \left(\frac {\gamma}2 \; N \sum_{ \sigma=0}^{q-1} [x (\sigma)]^2 
\right) \; \frac{N!}{\Pi_{\sigma=0}^{q-1} [N x(\sigma)]!}
 \nonumber \\
 &=& \int_0^{1\ (R)} \Pi _{\sigma=1}^{q-1} \; dx(\sigma) \;
 \exp \left( - N { f}[\{x (\sigma) \} ] \right)
\label{zpotts2}
\end{eqnarray}
to the leading order in $N$. The subscript $(R)$ indicates that the sum or
the integral must be restricted to the normalized subspace  
$\sum_{\sigma=0}^{q-1} x(\sigma)=1$. The ``free-energy'' density 
functional ${ f}$
appearing in (\ref{zpotts2}) is
\begin{equation}
{ f}[\{x (\sigma) \}]= \sum_{\sigma =0} ^{q-1} \bigg\{ - \frac{\gamma}2
 [x(\sigma)]^2 + x(\sigma ) \ln x (\sigma ) \bigg\} \qquad .
\label{fff}
\end{equation}
In the limit of large $N$, the integral in (\ref{zpotts2}) may be 
evaluated by the saddle-point method. The Potts free-energy (\ref{pottsfe})
then reads
\begin{equation}
f_{Potts} (q) =  \min_{\{x (\sigma ) \} } 
{ f}[\{x_\sigma\}]
\end{equation}
and the problem becomes that of analyzing the minima of $ f$.
Given the initial formulation of the problem, each possible value of
$\sigma$ among $0,\ldots , q-1$ plays the same role; indeed $f$ is
invariant under the permutation symmetry of the different
$q$ values.
However, we should keep in mind that such a symmetry could be 
broken by the minimum (see section \ref{sect_statisticalPhysics}). 
We shall see that
depending on the value of the connectivity $\gamma$, the permutation
symmetry may or may not be broken, leading to a phase transition in
the problem which coincides with the birth a giant component in the
associated random graph.

\subsubsection{Symmetric saddle-point.}

Consider first the symmetric extremum of $ f$,  
\begin{equation}
\label{symsad}
x ^{sym}(\sigma)=\frac 1q, \qquad \forall \sigma =0 , \ldots , q-1 .
\end{equation}
We have 
\begin{equation}
f ^{sym} _{Potts} (q) = - \ln q - \frac{\gamma}{2 q} \; \; \; .
\end{equation}
Taking the Legendre transform of this free-energy, see
(\ref{colle},\ref{numclus}), we get for the logarithm of the cluster
distribution density
\begin{equation}
\omega ^{sym} (c) = -\frac{\gamma}2 - (1-c) (1+\ln \gamma - \ln
[2 (1-c)]) \qquad .
\label{omsym}
\end{equation}
$\omega ^{sym} (c)$ is maximal and null at $c^{sym}(\gamma )=1-
\frac{\gamma}2$, a result that cannot be true for connectivities larger
than two and must break down somewhere below. Comparison with the rigorous
derivation in random graph theory 
indicates that the symmetric result is exact as long as $\gamma \le 
\gamma _c =1$ and is false above the percolation threshold
$\gamma _c$. The failure of the symmetric extremum in the presence
of a giant component proves the onset of symmetry breaking.

To understand the mechanism responsible for the symmetry breaking,
we look for the local stability of the symmetric saddle-point
(\ref{symsad}) and compute the eigenvalues of the Hessian matrix
\begin{equation}
{ M} _{\sigma , \tau } = \frac{ \partial ^2}{\partial x(\sigma )
x (\tau) } { f}[\{x (\sigma\})] \biggr| _{sym, (R)}
\qquad ,
\end{equation}
restricted to the normalized subspace. The simple algebraic structure
of ${ M}$ allows an exact computation of its $q-1$ eigenvalues for
a generic integer $q$. We find a non degenerate eigenvalue $\lambda _0
= q (q-\gamma )$ and another eigenvalue $\lambda _1 = q-\gamma$ with
multiplicty $q-2$. The analytic continuation of the eigenvalues to
real $q \to 1$ lead to the single value $\lambda = 1-\gamma$
which changes sign
at the percolation threshold $\gamma _c$.  Therefore, the symmetric
saddle-point is not a local minimum of ${ f}$ above 
$\gamma_c$, showing that a more complicated saddle-point has to be found.   

\subsubsection{Symmetry broken saddle-point.}

The simplest way to break the symmetry of the problem
is to look for solutions in which one among the $q$ values appears
more frequently than the others. Therefore we look for a saddle-point of
the form
\begin{eqnarray}
x (0) &=& \frac 1q [1+(1-q) s] \nonumber \\
x (\sigma) &=& \frac 1q [1-s] \ , \qquad  (\sigma=1,...,q-1).
\label{fracrot}
\end{eqnarray}
The symmetric case can be recovered in this enlarged subspace of 
solutions by setting $s=0$. The free-energy of the Potts model
is obtained by plugging the fractions
(\ref{fracrot}) into (\ref{fff}). In the limit $q\to 1$ of interest,
\begin{equation}
{ f}[\{x_\sigma\}]=-\frac{\gamma}2 + (q-1) f_{Potts}
(s,\gamma)+O((q-1)^2)
\end{equation}
with
\begin{equation}
f_{Potts} (s,\gamma)=\frac{\gamma}2 (1 - \frac 12 s^2) 
-1+s+(1-s) \ln(1-s)
\end{equation}
Minimization of $f_{Potts}(s,\gamma)$ with respect to the order parameter $s$
shows that for $\gamma \leq 1$ the symmetric solution $s=0$ is
recovered, whereas for $\gamma>1$ there exists a non vanishing optimal
value $s^*(\gamma)$ of $s$ that is solution of the implicit equation
\begin{equation}
1- s^* = \exp ( - \gamma \; s^* ) \qquad .
\label{resexa}
\end{equation}
The stability analysis (which we will not give here) shows that the solution
is stable for any value of $\gamma$. The interpretation of
$s^*(\gamma)$ is straightforward: $s^*$ is the fraction of vertices
belonging to the giant cluster. The average fraction of connected
components $c(\gamma )$ equals $-f_{Potts} (s^*(\gamma ),\gamma )$,
see (\ref{numclus}), in perfect agreement with exact results by 
Erd\"os and Renyi.

\subsection{Discussion.}

Further results on the properties of
random graphs can be extracted from the previous type of
calculation. We shall examine two of them.

\subsubsection{Scaling at the percolation point.}
\label{critexpo2}

Given the interpretation of $s^*(\gamma)$ for any large but finite
value of $N$, we may define the probability of existence of a cluster
containing $N s$ sites as follows
\begin{equation}
{ P}(s,N) \simeq \frac{\exp(N f(s,\gamma))}{\exp(N f(s^*,\gamma))}
\end{equation}
In the infinite size limit this leads to the expected result
\begin{equation}
\lim_{N \to \infty} { P}(s,N)=\delta(s-s^*(\gamma))
\end{equation}
In order to describe in detail how sharp (in $N$) the transition is at
$\gamma=1$, we need to consider corrections to the saddle point
solutions by making an expansion of the free-energy $f_{Potts}
(s,\gamma =1)$ in the order parameter $s$.  At threshold, we have 
$s^*(1)=0$ and $f_{Potts} (s,1)=-s^3/6+O(s^4)$ and therefore
\begin{equation}
{ P}(s,N) \simeq \exp(-N s^3/6)
\end{equation}
In order to keep the probability finite at the critical point the only
possible scaling for $s$ is $s=O(N^{-1/3})$ which leads to a size of
the giant component at criticality $N \times N^{-1/3}=N^{2/3}$, in
agreement with the Erd\"os-R\'enyi results.

\subsubsection{Large deviations.}

The knowledge of the Potts free-energy for any value of $q$ allows one to
compute its Legendre transform, $\omega (c)$. The computation
does not show any difficulty and we do not reproduce the results
here~\cite{Morgante}. 
Phase transitions are also found to take place for rare events (graphs
that do not dominate the cluster probability distribution). 
Notice that we consider here random graphs obtained by deleting edges from
$K_N$ with a fixed probability. Large deviations results indeed depend
strongly on the process of generating graphs.

As a typical example of what can be found using statistical mechanics,
let us mention this simple result
\begin{equation}
\omega (c=1 ) = - \frac {\gamma}2
\qquad ,\label{poiu}
\end{equation}
for all connectivities $\gamma$. The above identity means that the
probability that a random graph has $N - o(N)$ connected components
decreases as $\exp(-\gamma N/2)$ when $N$ gets large. This result
may be easily understood. Consider for instance graphs with $\gamma N$
edges made of a complete graph on $\sqrt{ 2 \gamma N}$ vertices plus
$N-\sqrt{ 2 \gamma N}$ isolated vertices. The fraction of connected 
components in
this graph is $c=1-O(1/\sqrt N) \to 1$. The number of such graphs
is simply the number of choices of $\sqrt{ 2 \gamma N}$ vertices 
among $N$ ones. Taking into account the edge deletion probability
$1-p=1-\gamma/N$, one easily recovers (\ref{poiu}).

\subsubsection{Conclusion.}

The random graph problem is a nice starting point to test ideas and
techniques from statistical mechanics. First, rigorous results are
known and can be confronted to the outputs of the
calculation. Secondly, analytical calculations are not too difficult
and can be exploited easily.

As its main focus, this section aimed at exemplifying the strategy
used in more complicated, e.g. K-Satisfiability, problems. The
procedure of analytic continuation, which is at the root of the replica
approach, appears nicely in the computation of the Potts free-energy
and is shown to give exact results (though in a non rigorous way).
The power of the approach is impressive. Many quantities can be 
computed and rather subtle effects such as large deviations are easily
obtained in a unique framework.

At the same time, the main weakness of the statistical mechanics
approach is also visible. Most interesting effects are obtained when
an underlying symmetry is broken. But the structure of the broken
saddle-point subspace is far from obvious, in contrast to the Ising
case of the previous section. There is at first sight some kind of
arbitrariness in the search of a saddle-point of the form of
(\ref{fracrot}). In the absence of a well-established and rigorous
procedure, the symmetry breaking schemes to be used must satisfy at
least basic self-consistency checks (plausibility of results, local
stability, ...). In addition, theoretical physicists have developed
various schemes that are known to be efficient for various classes of
problems but (fail in other cases). A kind of standard lore, of
precious help to solve new problems, exists and is still waiting for
firm mathematical foundations.

\section{Random K-satisfiability problem}

\label{sect_ksat}

In what follows we shall describe the main steps of the replica
approach to the statistical mechanics analysis of the Satisfiability
problem. The interested reader may find additional details concerning
the calculations in several published
papers~\cite{MonassonZecchina96,MonassonZecchina97,MonassonZecchina98,MZKST1999,MZKST_RSA,Monasson_diluiti,Variational}
and in the references therein.

The satisfaction of constrained Boolean formulae is a key issue in
complexity theory. Many computational problems are known to be
NP-complete \cite{GareyJohnson79,Papadimitriou} through a polynomial
mapping onto the K-Satisfiability (SAT) problem, which in turn was the
first problem shown to be NP-complete by Cook in 1971
\cite{Cook}.

Recently~\cite{Hogg}, there has been much interest in a random version
of the K-SAT problem defined as follows. Consider $N$ Boolean
variables $x_i$, $i=1,\ldots ,N$. Call a clause $C$ the logical OR of
$K$ randomly chosen variables, each of them being negated or left
unchanged with equal probabilities.  Then repeat this process by
drawing independently $M$ random clauses $C_\ell$, $\ell =1 ,\ldots
,M$. The logical AND of all these clauses is a ``formula'', referred
to as ${ F}$. It is said to be satisfiable if there exists a logical
assignment of the $x$s evaluating ${ F}$ to true, and unsatisfiable
otherwise.

Numerical experiments have concentrated on the study of the
probability $P_N (\alpha , K)$ that a randomly chosen ${ F}$ having
$M=\alpha N$ clauses be satisfiable. For large sizes, a remarkable
behaviour arises: $P_N$ seems to reach unity for $\alpha < \alpha _c
(K)$ and vanishes for $\alpha > \alpha _c (K)$ when $N \to
\infty$~\cite{hardness,Hogg}.  Such an abrupt threshold behaviour,
separating a SAT phase from an UNSAT one, has indeed been rigorously
confirmed for 2-SAT, which is in P, with $\alpha _c (2) =1$
\cite{2SAT_1,2SAT_2}.  For larger $K \ge 3$, K-SAT is NP-complete and
much less is known.  The existence of a sharp transition has not been
rigorously proved but estimates of the thresholds have been found~:
$\alpha _c (3) \simeq 4.3$
\cite{KirkpatrickSelman}.  Moreover, some rigorous lower and upper
bounds to $\alpha _c (3)$ (if it exists), $\alpha _{l.b.}=3.14$ and
$\alpha _{u.b.} = 4.51$ respectively have been established (see the
review articles dedicated to upper and lower bounds contained in this
TCS special issue).

The interest in random K-SAT arises partly from the following fact: it
has been observed numerically that hard random instances are generated
when the problems are critically constrained, i.e., close to the
SAT/UNSAT phase boundary \cite{hardness,Hogg}. The study of such hard
instances represents a theoretical challenge towards an understanding
of complexity and the analysis of exact algorithms. Moreover, hard
random instances are also a test-bed for the optimization of heuristic
(incomplete) search procedures, which are widely used in practice.

Statistical mechanics provides new intuition on the nature of the
solutions of random K-SAT (or MAX-K-SAT) through the introduction of
an order parameter which describes the geometrical structure of the
space of solutions. In addition, it gives also a global picture of the
dynamical operation of search procedures and the computational
complexity of K-SAT solving.

\subsection{K-SAT energy and the partition function.}

To apply the statistical physics approach exemplified on the random
graph problem, one has to identify the energy function corresponding
to the K-SAT problem.

The logical values of an $x_i$ can be represented by a binary variable
$S_i$, called a spin, through the one-to-one mapping $S_i=-1$
(respectively $+1$) if $x_i$ is false (resp. true). The random clauses
can then be encoded into an $M\times N$ matrix $C_{\ell i}$ in the
following way~: $C_{\ell i}=-1$ (respectively $+1$) if the clause
$C_\ell $ includes $\overline{x_i }$ (resp. $x_i$), $C_{\ell i}=0$
otherwise. It can be checked easily that $\sum _{i=1}^N C_{\ell i}
S_i$ equals the number of wrong literals in clause $\ell$. Consider
now the cost-function $E[{\bf C } , {\bf S }]$ defined as the number
of clauses that are not satisfied by the logical assignment
corresponding to configuration ${\bf S}$.

\begin{equation}
E[{\bf C},{\bf S}]=  
\sum _{\ell =1}^{M} \delta \left(
\sum _{i=1}^N C_{\ell i} S_i  +K \right) \quad ,
\label{cost}
\end{equation}
where $\delta (j)=1$ if $j=0$, zero otherwise, denotes the Kronecker
function.  The minimum (or ground state -GS) $E[{\bf C}]$ of $E[{\bf C
} , {\bf S }]$, is the lowest number of violated clauses that can be
achieved by the best possible logical assignment
\cite{MonassonZecchina97}. 
 $E[{\bf C}]$ is a random variable that becomes highly concentrated
around its average value $E_{GS} \equiv
\overline{E[{\bf C}] }$ in the large size limit \cite{Self_averaging}. The
latter is accessible through the knowledge of the averaged logarithm
of the generating function
\begin{equation}
Z[{\bf C } ]= \sum _{\bf S} \exp \left( - E[{\bf C } , {\bf S }] / T \right)
\label{partfunc}
\end{equation}
since 
\begin{equation}
E_{GS}  = - T \;\overline  {\log Z[{\bf C } ] } + O(T^2) \qquad,
\label{partfunc2}
\end{equation}
when the auxiliary parameter $T$ is sent to zero. Being the minimal
number of violated clauses, $E_{GS}$ equals zero in the sat region and
is strictly positive in the unsat phase. The knowledge of $E_{GS}$ as
a function of $\alpha$ therefore determines the threshold ratio
$\alpha _c(K)$.

\subsection{The average over the disorder.}

The calculation of the average value of the logarithm of the partition
function in (\ref{partfunc2}) is an awkward one. To circumvent this
difficulty, we compute the $n^{th}$ moment of $Z$ for integer-valued
$n$ and perform an analytic continuation to real $n$ to exploit the
identity $\overline {Z[{\bf C } ] ^ n } = 1 + n \overline{\log Z[{\bf
C } ] } + O(n^2)$. The $n^{th}$ moment of $Z$ is obtained by
replicating $n$ times the sum over the spin configurations ${\bf S}$
and averaging over the clause distribution \cite{MonassonZecchina97}
\begin{equation}
\overline  {Z[{\bf C } ] ^ n } = \sum _{{\bf S}^1 , {\bf S}^2 , \ldots
,{\bf S}^n} \overline 
{\exp \left( -  \sum _{a=1} ^n E[{\bf C } , {\bf S }^a] / T \right) } \quad ,
\label{nthmoment}
\end{equation}
which in turn may be viewed as a generating function in the variable
$e^{-1/T}$.

In order to compute the expectation values that appear in
eq.(\ref{nthmoment}), one notices that each individual term
\begin{equation}
z [\{ {\bf S}^a \} ] = 
\overline{ \exp \left( -\frac{1}{T} \sum _{a=1} ^n
E[ {\bf C } , {\bf S }^a] \right) }
\end{equation}
factorises over the sets of different clauses due to the absence of
any correlation in their probability distribution. It follows
\begin{equation}
z [ \{ {\bf S}^a \}] = 
\left( \zeta _K [ \{ {\bf S}^a \} ] \right) ^{M}
\qquad ,
\end{equation}
where each factor is defined by
\begin{equation}
\zeta _K [ \{ {\bf S}^a \} ] = \overline{ \exp \left[ -\frac{1}{T} \sum _{a=1} ^n
\delta \left( \sum _{i=1}^N C_i S_i^a + K \right) \right] }
\qquad , \label{appe}
\end{equation}
with the bar denoting the uniform average over the set of $2^K {N
\choose K}$ vectors of $N$ components $C_i =0 ,\pm 1$ and of squared
norm equal to $K$.

Resorting to the identity,
\begin{equation}
\delta \left( \sum _{i=1}^N C_i S_i^a + K \right) =
\prod _{i / C_i \ne 0} \delta \left( S_i^a  + C _i \right) \qquad,
\end{equation}
one may carry out the average over in disorder in eq.(\ref{appe}) to obtain
\begin{equation}
\zeta _K [ \{ {\bf S}^a \} ] = \frac 1{2^K} \sum _{C_1 ,\ldots , C_K =\pm 1} \frac 1{N^K}
\sum _{i_1 ,\ldots , i_K =1} ^N \exp \left\{ -\frac{1}{T} \sum _{a=1} ^n
\prod _{\ell =1}^K \delta \left[ S_{i_\ell} ^a + C _\ell\right] \right\}
\end{equation}
up to negligible $O(1/N)$ contributions. 

The averaged term in the r.h.s. of (\ref{nthmoment}) depends on the
$n\times N$ spin values only through the $2^n$ occupation fractions
$x({\vec \sigma} )$ labeled by the vectors ${\vec \sigma}$ with $n$
binary components; $x({\vec \sigma} )$ equals the number (divided by
$N$) of labels $i$ such that $S_i^a=\sigma ^a$, $\forall a=1,\ldots
,n$. It follows that $\zeta _K [ \{ {\bf S}^a \}] =\zeta _K [ {\bf x}
]$ where
\begin{eqnarray}
\zeta _K [{\bf  x} ] = \frac 1{2^K} \sum _{C_1 ,\ldots , C_K =\pm 1} &&
\sum _{\vec \sigma _1 , \ldots , \vec \sigma _K}  
x (- C_1 \;\vec \sigma _1  ) \ldots 
x (- C_K\; \vec \sigma _K ) \times \\ \nonumber
&& \exp \left\{ 
-\frac{1}{T} \sum _{a=1} ^n \prod _{\ell =1}^K \delta \left[ \sigma _\ell
 ^a - 1 \right] \right\}
\quad .
\end{eqnarray}

To leading order in $N$ (e.g., by resorting to a saddle point
integration), the final expression of the $n^{th}$ moment of $Z$ can
be written as $\overline {Z[{\bf C } ] ^n } \simeq \exp ( -N \;
f_{opt}/T)$ where $f_{opt}$ is the optimum (in fact the minimum for
integer $n$) over all possible ${\bf x}$s of the functional
\cite{MonassonZecchina97}
\begin{equation}
f[ {\bf x } ] = e[{\bf x}] + \frac 1T  
\sum _{\vec \sigma} x({\vec \sigma}) \log x({\vec \sigma}) \qquad ,
\label{eqfunc}
\end{equation}
with
\begin{equation}
e[{\bf x }] =
\alpha \ln \left[ \sum _{\vec \sigma_1 ,... , \vec \sigma_K } x
(\vec \sigma_1 ) ... x (\vec \sigma_K ) \ \exp \left( -
\frac{1}{T} \sum _{a=1} ^n \prod _{\ell =1}^K \delta [ \sigma_{\ell}^a - 1]  
\right) \right] \ .
\label{expre}
\end{equation}
Note the similarities between equations (\ref{eqfunc}) and
(\ref{fff}).  While in the random graph or Potts model case $\sigma$
took on $q$ values, the K-SAT model requires the introduction of $2^n$
vectors $\vec \sigma$. In both cases, an analytic continuation of the
free-energy to non integer values of $q$ or $n$ has to be performed.
Finally, note that the optimum of $f$ fulfills $x (\vec \sigma ) = x
(-\vec \sigma ) $ due to the uniform distribution of the disorder $C$.

\subsection{Order parameter and replica-symmetric saddle-point equations.}

The optimization conditions over $f[ {\bf x} ]$ provide $2^n$ coupled
equations for the ${\bf x}$s. Notice that $f$ is a symmetric
functional, invariant under any permutation of the replicas $a$, as is
evident from equation (\ref{nthmoment}). An extremum may thus be
sought in the so-called replica symmetric (RS) subspace of dimension
$n+1$ where $x({\vec \sigma} )$ is left unchanged under the action of
the symmetric group. In the limit of interest, $T \to 0$, and within
the RS subspace, the occupation fractions may be conveniently
expressed as the moments of a probability density $P(m)$ over the
range $-1\le m\le 1$
\cite{MonassonZecchina97},
\begin{equation} \label{defpdem}
x  (\sigma _1 ,\sigma _2 ,\ldots ,\sigma _n ) =
\int _{-1} ^1 dm \; P(m) \prod _{a=1} ^n
\left( \frac{1+m \sigma ^a} {2} \right) \qquad .
\end{equation}
$P(m)$ is not uniquely defined by (\ref{defpdem}) for integer values
of $n$ but acquires some precise meaning in the $n\to 0$ limit. It is
the probability density of the expectation values of the spin
variables over the set of ground states.  Consider a formula ${ F}$
and all the spin configurations ${\bf S}^{(j)}$, $j=1,\ldots ,{Q}$
realizing the minimum $E[{\bf C } ]$ of the cost-function $E[{\bf C
},{\bf S }]$, that is the solutions of the MAX-SAT problem defined by
$F$. Then define the average magnetizations of the spins
\begin{equation}
m_i = \frac 1{Q} \sum _{j=1} ^{Q} S_i^{(j)} \quad ,
\end{equation}
over the set of optimal configurations. Call $H({\bf C },m)$ the
histogram of the $m_i$s and $H(m)$ its quenched average, i.e., the
average of $H({\bf C },m)$ over the random choices of the formulae ${
F}$. $H(m)$ is a probability density over the interval $-1\le m\le 1$
giving information on the distribution of the variables induced by the
constraint of satisfying all the clauses. In the absence of clauses,
all assignments are solutions and all magnetizations vanish:
$H(m)=\delta(m)$ and variables are not constrained. Oppositely,
variables that always take the same value in all solutions, if any,
have magnetizations equal to $+1$ (or $-1$): such variables are
totally constrained by the clauses.

As discussed in ref.~\cite{MonassonZecchina97}, if the RS solution is
the global optimum of (\ref{eqfunc}) then $H(m)$ equals the above
mentioned $P(m)$ in the limit of large sizes $N\to \infty$. Therefore,
the order parameter arising in the replica calculation reflects the
``microscopic'' structure of the solutions of the K-SAT problem.

At this stage of the analysis it is possible to perform the analytic
continuation $n \to 0$ since all the functionals have been expressed
in term of the generic number of replicas $n$. Such a process leads to
a self-consistent functional equation for the order parameter $P(m)$,
which reads
\begin{eqnarray}
P(m) &=&\frac 1{1-m^2}\int_{-\infty }^\infty du\ \cos \left[ \frac u2\ln
\left( \frac{1+m}{1-m}\right) \right] \times  \nonumber \\
&& \exp \left[ -\alpha K+\alpha K\int_{-1}^1\prod_{\ell =1}^{K-1}dm_\ell
P(m_\ell )\cos \left( \frac u2\ln A_{(K-1)} \right) \right]  
\label{pxint}
\end{eqnarray}
with 
\begin{equation}
A _{(K-1)} \equiv A _{(K-1)} (\left\{ m_\ell \right\} ,\beta)= 1+(e^{-\beta
}-1)\prod_{\ell =1}^{K-1}\left( \frac{1+m_\ell }2\right) \; \; \;,
\label{defA}
\end{equation}
and $\beta \equiv 1/T$. The corresponding replica symmetric free--energy 
density reads 
\begin{eqnarray} \label{f56}
- \beta \, f_{opt} (\alpha , T) &=& \ln 2 + 
\alpha (1-K)\int_{-1}^1\prod_{\ell =1}^K dm_\ell
P(m_\ell )\ln A_{(K)}  \nonumber \\
&+&\frac{\alpha K}2\int_{-1}^1\prod_{\ell =1}^{K-1}dm_\ell P(m_\ell )\ln
A_{(K-1)} \nonumber \\
&-& \frac 12\int_{-1}^1dmP(m)\ln (1-m^2) \qquad .  \label{Frs}
\end{eqnarray}
It can be checked that equation (\ref{pxint}) is recovered when
optimizing the free-energy functional (\ref{f56}) over all (even)
probability densities $P(m)$ on the interval [-1,1].

\subsection{The simple case of K=1.}

Before entering in the analysis of the saddle-point equations for
general $K$, it is worth considering the simple $K=1$ case which can
be solved either by a direct combinatorial method or within the
statistical mechanics approach. Though random 1-SAT does not present
any critical behaviour (for finite $\alpha$), its study allows an
intuitive understanding of the meaning and correctness of the
statistical mechanics approach.

For $K=1$, a sample of $M$ clauses can be defined completely by giving
directly the numbers $t_i$ and $f_i$ of clauses imposing that a
certain Boolean variable $S_i$ must be true or false respectively.
The partition function corresponding to a given sample reads
\begin{equation}
Z[\{ t, f\}]=\prod _{i=1}^N (e^{-\beta t_i} +e^{-\beta f_i}) \; \; \;,
\end{equation}
and the average over the disorder gives 
\begin{eqnarray}
\frac 1N\overline{\ln Z[\{ t, f\}]} &=& \frac 1N \sum _{\{t_i,f_i\}} \frac{M!
}{\prod_{i=1}^N (t_i! f_i!)} \ln Z[\{ t, f\}]  \nonumber \\
&=& \ln 2 - \frac{\alpha \beta}{2} + \sum _{l=-\infty}^{\infty} e^{-\alpha}
I_l (\alpha)\; \ln \left(\cosh \left(\frac{\beta l}{2}\right)\right) \; \;
\;,
\end{eqnarray}
where $I_l$ denotes the $l^{th}$ modified Bessel function. The zero
temperature limit gives the ground state energy density
\begin{equation}
e_{GS}(\alpha )=\frac{\alpha}{2} [ 1-e^{-\alpha } I_0(\alpha ) - e^{-\alpha
} I_1(\alpha )]
\end{equation}
and the ground state entropy density
\begin{equation}
s_{GS}(\alpha )= e^{-\alpha } I_0(\alpha )\; \ln 2 \; \; \;.
\end{equation}

For any $\alpha>0$, the ground--state energy density is positive and
therefore the overall Boolean formula is false with probability one.
Also, the entropy density is finite, {\em i.e.}, the number of minima
of the energy for any $\alpha$ is exponentially large.  Such a result
can be understood by noticing that there exist a fraction of
unconstrained variables $e^{-\alpha } I_0(\alpha )$ which are subject
to equal but opposite constraints $t_i=f_i$.

The above results are recovered in the statistical mechanics
framework, thereby showing that the RS Ansatz is exact for all $\beta$
and $\alpha$ when $K=1$.

The solution of the saddle-point equation (\ref{pxint}) can be found
for any temperature $T$ leading to the expression
\begin{equation}
P(m)= \sum _{\ell=-\infty}^{\infty} e^{-\alpha} I_\ell (\alpha) \; \delta
\left( m - \tanh \left( \frac{\beta \ell}{2} \right) \right) \; \; \; .
\label{Saddlek=1}
\end{equation}
In the limit of interest $\beta \to \infty$, this
formula reads 
\begin{equation}
P(m)=e^{-\alpha } I_0(\alpha ) \delta(m)+\frac 12 (1-e^{-\alpha } I_0(\alpha
)) \left(\delta(m-1)+\delta(m+1)\right) \; \;.  \label{resuk=1}
\end{equation}

\begin{figure}
\begin{center}
\resizebox{0.4\linewidth}{!}{\includegraphics{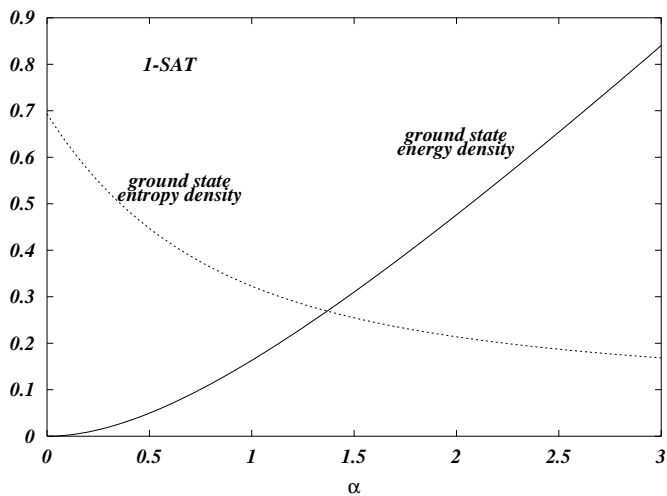}}
\end{center}
\caption{Energy density (bold line) and entropy
density (thin line) versus $\alpha$ in a random 1-SAT formula, in the limit $N
\to \infty$.}
\label{ksat_fig1}
\end{figure}

As shown in figure~\ref{ksat_fig1}, the fraction of unconstrained
variables is simply associated with the unfrozen spins and thus gives
the weight of the $\delta$--function at $m=0$.  On the contrary, the
non-zero value of the fraction of violated clauses, proportional to
the ground-state energy density, is due to the presence of completely
frozen (over constrained) spins of magnetizations $m=\pm 1$. Such a
feature remains valid for any $K$.

\subsection{Sat phase: structure of the space of solutions.}

We start by considering the sat phase. An interesting quantity to look
at is the typical number of solutions of the random K-SAT problem;
this quantity can be obtained from the ground state entropy density
$s_{GS}(\alpha)$ given by eq.(\ref{Frs}) in the $\beta\to\infty$
limit.

\begin{figure}
\begin{center}
\resizebox{0.5\linewidth}{!}{\includegraphics{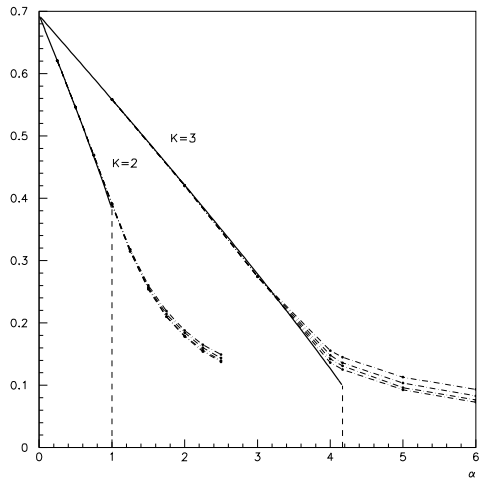}}
\end{center}
\caption{RS estimate for the entropy density in random 2-SAT and 3-SAT
below their thresholds.  RSB corrections due to clustering are absent
in 2-SAT and very small (within few a percent) in 3-SAT. The dots
represent the results of exact enumerations in small systems ($N$
ranging from $20$ to $30$, see ref.~\cite{MonassonZecchina96})}
\label{ksat_fig2}
\end{figure}

In the absence of any clauses, all assignments are solutions:
$s_{GS}(\alpha=0) =\ln 2$. We have computed the Taylor expansion of
$s_{GS}(\alpha)$ in the vicinity of $\alpha=0$, up to the seventh
order in $\alpha$.  Results are shown in Figure~\ref{ksat_fig2}.  It
is found that $s_{GS}(\alpha_c=1)= .38$ and $s_{GS}(\alpha =4.2)=.1$
for 2-SAT and 3-SAT respectively: just below threshold, solutions are
exponentially numerous. This result is confirmed by rigorous
work~\cite{Rigorous_entropy}.

More involved calculations, including replica symmetry breaking (RSB)
effects \cite{Variational}, have shown that the value of the entropy
is insensitive to RSB in the sat phase. Therefore the RS calculation
provides a quite precise estimate of the entropy (believed to be exact
at low $\alpha$ ratios, see Talagrand's paper in this volume for a
discussion).

\begin{figure}
\begin{center}
\resizebox{0.5\linewidth}{!}{\includegraphics{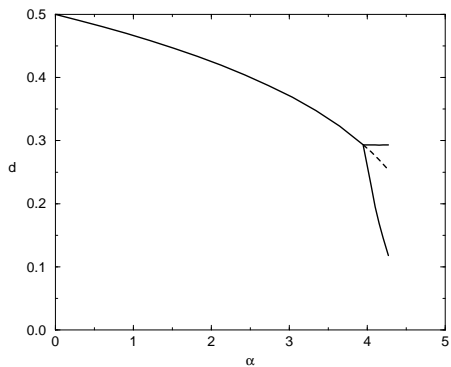}}
\end{center}
\caption{Variational RSB estimate for the clustering of solutions below 
$\alpha_c$ for $3-SAT$. $d$ is the typical Hamming distance between
solutions. The splitting of the curves at $\alpha\simeq 4$ corresponds
to clustering. There appear two characteristic distances, one within
each cluster and one between solutions belonging to different clusters.}
\label{ksat_fig3}
\end{figure}

Recent analytical calculations for 3-SAT \cite{Variational} (also
confirmed by numerical investigations) indicate that the RS theory
breaks down at a definite ratio $\alpha_{RSB}$ below $\alpha_c$, where
the solutions start to be organized into distinct clusters. The
meaning of this statement is as follows. Think of the space of spins
configurations as the $N$-dimensional hypercube. Optimal assignments
are a subset of the set of $2^N$ vertices on the hypercube. Replica
symmetry amounts to assuming that any pair of vertices are
a.s. separated by the same Hamming distance $d$, defined as the
fraction of distinct spins in the corresponding configurations. In
other words, solutions are gathered in a single cluster, of diameter
$d\,N$. RSB variational calculations
\cite{Variational} show that this simplifying assumption is not
generally true in the whole sat phase and that another scenario may
take place close to threshold:
\begin{itemize}
\item Below $\alpha_{RSB}$ the space of solutions
is replica symmetric.  There exist one cluster of solutions
characterized by a single probability distribution of local
magnetizations. The Hamming distance $d$ is a decreasing function
of $\alpha$, starting at $d(0)=1/2$.  
\item At $\alpha_{RSB} \simeq 4.0$, the space of solutions breaks into 
a large number (polynomial in $N$) of different clusters. Each cluster
contains an exponential number of solutions. The typical Hamming
distance $d_0$ between solutions belonging to different clusters is
close to $0.3$ and remains nearly constant (it is slightly decreasing)
up to $\alpha_c$, indicating that the centers of these clusters do not
move on the hypercube when more and more clauses are added. Within
each cluster, solutions tend to become more and more similar, with a
rapidly decreasing intra-cluster Hamming distance $d_1$.
\end{itemize}
Figure~\ref{ksat_fig3} provides a
qualitative representation of the clustering process.
The fact that the Hamming distance can take two values at most is a
direct consequence of the RSB Ansatz. In reality, the distance
distribution could be more complicated. The key point is that
statistical mechanics calculations strongly support the idea that the
space of solutions has a highly organized structure, even in the sat
phase.

Recently, the exact solution of the balanced version of random K-SAT
\cite{RicciWeigtZecchina} has provided a concrete example in which
the appearance of clustering before the sat/unsat transition can be
studied both analytical and numerically. Note that this phenomenon is
strongly reminiscent of what happens in some formal multi-layer neural
networks models~\cite{Engel}.

\subsection{Unsat phase: the backbone and the order of the phase transition.}

In the unsat phase, it is expected that $O(N)$ variables become
totally constrained, {\em i.e.} take on the same value in all the
ground states. Such a hypothesis, which of course needs to be verified
{\em a posteriori}, corresponds to a structural change in the
probability distribution $P(m)$ which develops Dirac peaks at $m=\pm
1$.

In the limit of interest ($T \to 0$), to describe the accumulation of
the magnetization on the borders of its domain ($m\in [-1;1]$), we
introduce the rescaled variable $z$, implicitly defined by the
relation $m=\tanh (z/T)$, see equation (\ref{resuk=1}). Calling $R(z)$
the probability density of the $z$s, the saddle-point equations read
\begin{eqnarray}
\label{Saddler}
R(z)&=&\int_{-\infty}^{\infty}\frac{du}{2\pi}\cos(uz) \exp \left[ -\frac{
\alpha K}{2^{K-1}} + \alpha K \times \right.\\ \nonumber
&&\left. \int _0 ^{\infty}\prod_{\ell=1}^{K-1} dz_\ell
R(z_\ell)\cos( u\ \hbox{\rm min} (1,z_1,\ldots,z_{K-1}))\right] \; \;.
\end{eqnarray}
The corresponding ground state energy density reads, see (\ref{Frs}), 
\begin{eqnarray}
e_{GS} (\alpha ) &=& \alpha (1-K) \int _0 ^{\infty} \prod _{\ell =1} ^K
dz_\ell R(z_\ell ) \hbox{min} (1,z_1,\ldots ,z_K)   \nonumber \\
&+& \frac{\alpha K}{2} \int _0 ^{\infty} \prod _{\ell =1} ^{K-1} dz_\ell
R(z_\ell ) \hbox{min} (1,z_1,\ldots ,z_{K-1}) - \int _0 ^{\infty} dz R(z) z
\ \ .  \label{energywithr}
\end{eqnarray}

It is easy to see that the saddle--point equation (\ref{Saddler}) is
in fact a self--consistent identity for $R(z)$ in the range $z\in
[0,1]$ only.  Outside this interval, equation (\ref{Saddler}) is
merely a definition of the functional order parameter $R$.

As discussed in detail in ref. \cite{MonassonZecchina97}, equations
(\ref{Saddler}) admit an infinite sequence of more and more structured
exact solutions of the form
\begin{equation}
R(z)=\sum _{l=-\infty} ^\infty r_\ell \ \delta \left( z - \frac {\ell}{q}
\right) \; \; \; ,  \label{rdep}
\end{equation}
having exactly $q$ peaks in the interval $[0,1[$, whose centers are
$z_\ell = \frac {\ell}{q}$, $\ell=0,\ldots ,q-1$.  The corresponding
energy density reads, from (\ref{rdep}) and (\ref{energywithr}),
\begin{eqnarray}
e_{GS} &=& \frac{\alpha (1-K)}{q} \left[ \left(\frac{1-r_0}{2}\right) ^K +
\sum _{j=1}^{q-1} \left(\frac{1-r_0}{2} - \sum _{l=1}^j r_l\right) ^K
\right]   \nonumber \\ &+& 
\frac{\alpha K}{2 q}\! \! \left[ \left(\frac{1-r_0}{2}\right) ^{K-1} + \sum
_{j=1}^{q-1} \left(\frac{1-r_0}{2} - \sum _{l=1}^j r_l\right) ^{K-1} \right]
\nonumber \\
&-& \sum _{j=1}^q \frac{j}{q} \gamma _j \left( \frac{r_0}{2} + 
\frac{r_j}{2} + \sum _{l=1}^{j-1} r_l \right) \qquad .  \label{enercas}
\end{eqnarray}
Though there might be continuous solutions to (\ref{Saddler}), it is
hoped that the energy of ground state can be arbitrarily well
approximated by the above large $q$ solutions.

\begin{figure}
\begin{center}
\resizebox{0.45\linewidth}{!}{\includegraphics{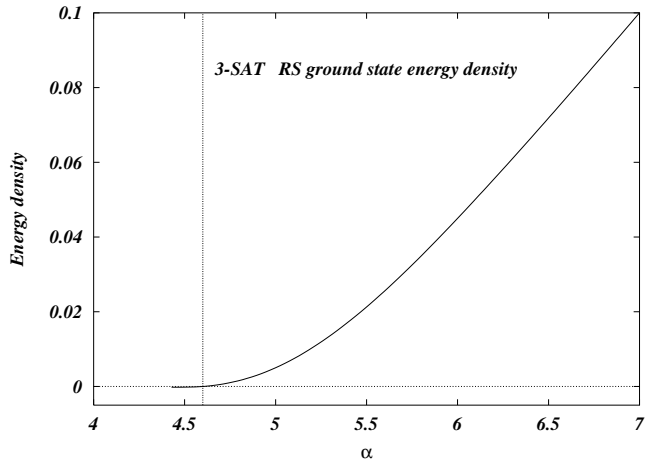}}
\end{center}
\caption{RS estimate for the ground state energy density,
{\i.e.}, the number of violated clauses divided by
$N$ in random 
3-SAT. The prediction is given as a function of $\alpha$, 
for $q\gg 1$ and in the limit 
$N \to \infty$. See ref.~\cite{MonassonZecchina97} for details.}
\label{ksat_fig4}
\end{figure}

The location of the sat/unsat threshold can be obtained for any $K$ by
looking at the value of $\alpha$ beyond which the ground state energy
becomes positive. For $2-SAT$ the exact result $\alpha_c(2)=1$ is
recovered whereas for $K>2$ the RS energy becomes positive at a value
of $\alpha$ (e.g., $\alpha_c(3) \simeq 4.6$ as shown in
figure~\ref{ksat_fig4}) which is sightly higher than the value
estimated by numerical simulations.

\subsubsection{A hint at replica symmetry breaking.}

The RS theory provides an upper bound for the thresholds for any
$K>2$, whereas the exact values can be obtained only by adopting a
more general functional form for the solution of the saddle-point
equations which explicitly breaks the symmetry between replicas (see
ref. \cite{Monasson_diluiti} for a precise discussion).  Such an issue
is indeed a relevant, and largely open, problem in the statistical
physics of random
systems~\cite{BanavarSherrington87,MezardParisi87c,VianaBray85,KanterSompolinsky87,DeDomGold,MottDeDom,DeDomMott,GoldLai}.

The general structure of the functional order parameter which
describes solutions that break the permutational symmetry among
replicas consists of a distribution of probability densities: each
Boolean variable fluctuates from one cluster of solutions to another,
leading to a site dependent probability density of local Boolean
magnetizations.  The distribution over all different variables then
provides a probability distribution of probability distributions.  The
above scheme can in principle be iterated, leading to more and more
refined levels of clustering of solutions. Such a scenario would
correspond to the so-called continuous RSB scheme
\cite{MezardParisiVirasoro}. However the first step solution could
suffice to capture the exact solution of random K-SAT, as happens in
other similar random systems
\cite{MezardParisiVirasoro}.

\subsubsection{Abrupt vs. smooth phase transition.}

\begin{figure}
\begin{center}
\resizebox{0.35\linewidth}{!}{\includegraphics{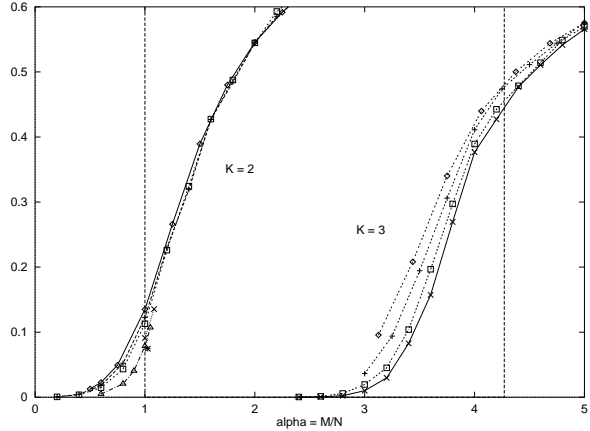}}
\end{center}
\caption{Numerical estimates of the value of the backbone order
parameter in 2-SAT and 3-SAT. The curves \cite{MZKST1999} are obtained
by complete enumerations in small systems (up to $N=500$
variables for 2-SAT and $N=30$ for 3-SAT) averaged over many samples.}
\label{ksat_fig5}
\end{figure}

Of particular interest are the fully constrained variables -- the so
called {\it backbone} component --, that is the $x_i$s such that $m_i
= \pm 1$. Within the RS Ansatz, the fraction of fully constrained
variables $\gamma (\alpha , K)$ can be directly computed from the
saddle-point equations.  Clearly, $\gamma (\alpha , K)$ vanishes in
the SAT region otherwise the addition of $\epsilon N$ new clauses to
${ F}$ would lead to a contradiction with a finite probability for any
$\epsilon > 0$.  Two kinds of scenarii have been found when entering
the unsat phase. For 2-SAT, $\gamma (\alpha , 2)$ smoothly increases
above the threshold $\alpha_c(2)=1$. For 3-SAT (and more generally
$K\ge 3$), $\gamma (\alpha , 3)$ exhibits a discontinuous jump to a
finite value $\gamma _c$ slightly above the threshold.  A finite
fraction of variables become suddenly over constrained when crossing
the threshold!  Numerical results on the growth of the backbone order
parameter are given in figure~\ref{ksat_fig5}.

\subsubsection{The random 2+p-SAT model.}

\begin{figure}
\begin{center}
\resizebox{0.5\linewidth}{!}{\includegraphics{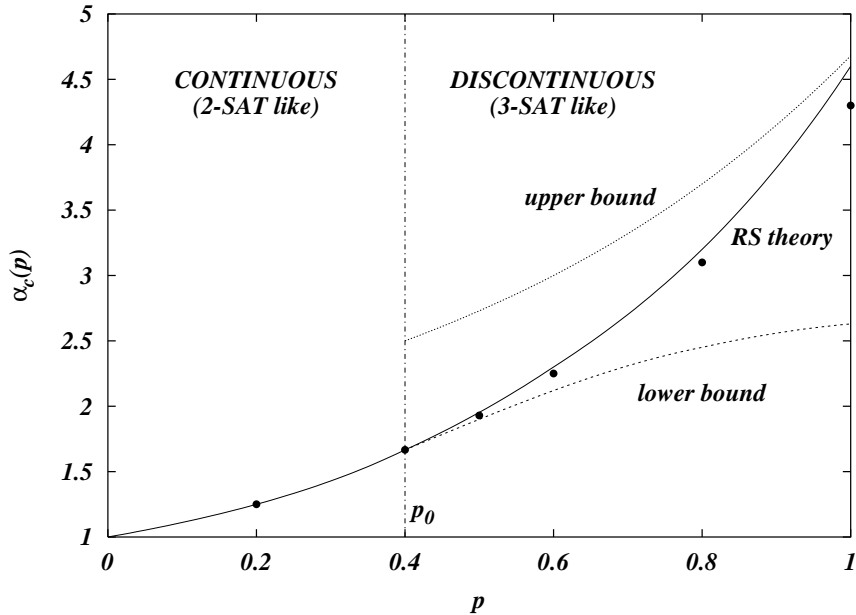}}
\end{center}
\caption{$\alpha_c(p)$ versus p in random 2+p-SAT.
Up to $p_0 \simeq .4$ $\alpha_c(p)=1/(1-p)$, in agreement with
rigorous results. For $p>p_0$ the transition becomes discontinuous in
the backbone order parameter and the RS theory provides an upper bound
for $\alpha_c(p)$ which is within a few percent of the results of
numerical simulations (dots) \cite{MZKST1999,MZKST_RSA}.}
\label{ksat_fig6}
\end{figure}

The sat/unsat transition is accompanied by a smooth (respectively
abrupt) change in the backbone component and therefore in the
structure of the solutions of the 2-SAT (resp. 3-SAT) problem. A
better way to understand how such a change takes place is to consider
a mixed model, which continuously interpolates between 2-SAT and
3-SAT.  The so-called $2+p$-SAT model \cite{MZKST1999} includes a
fraction $p$ (resp. $1-p$) of clauses of length two (resp. three).
2-SAT is recovered for $p=0$ and 3-SAT when $p=1$. The RS theory
predicts that, at the sat/unsat transition, the appearance of the
backbone component becomes abrupt when $p > p_0
\simeq 0.4$ (see figure~\ref{ksat_fig6}). On the contrary, when $p<p_0$,
the transition is smooth as in the 2-SAT case.  Such a scenario is
consistent with both rigorous results (see the paper by Achlioptas et
al. in this volume) based on the probabilistic analysis of simple
algorithm and with variational calculations \cite{Variational} which
include RSB effects.

An additional argument in favor of the above picture is given by the
analysis of the finite-size effects on $P_N (\alpha , K)$ and the
emergence of some universality for $p<p_0$. (The definition of $P_N$
was given when we began discussing the properties of K-SAT.) A
detailed account of these findings may be found
in~\cite{MZKST1999,MZKST_RSA}.  For $p<p_0$ the size of the critical
window where the transition takes place is observed to remain constant
and close to the value expected for 2-SAT. The critical behaviour is
the same as for the percolation transition in random graphs (see also
ref. \cite{ScalingWindow}). For $p>p_0$ the size of the window shrinks
following some non-universal exponents toward its statistical lower
bound~\cite{Wilson} but numerical data do not allow for any precise
estimate. The balanced version of $2+p$-SAT can be studied exactly and
both the phase diagram and the critical exponents turn out to behave
very similarly to the ones of $2+p$-SAT \cite{LeoneRicciZecchina}.

As we shall conclude in the next section, the knowledge of the phase
diagram of the 2+p-SAT model is very precious to understand the
computational complexity of 3-SAT solving.

\subsection{Computational complexity and dynamics.}
\label{subsect_ccd}

Numerical experiments have shown that the typical solving time of
search algorithms displays an easy-hard-easy pattern as a function of
$\alpha$ with a peak of complexity close to the threshold.  Since
computational complexity is strongly affected by the presence of a
phase transition, it is appropriate to ask whether the nature of this
phase transition plays an important role too. The peak in the search
cost seems indeed to scale polynomially with $N$ (even using
Davis-Putnam-like procedures) for the 2-SAT problem, where the
transition is continuous, and exponentially with $N$ in the 3-SAT
case, for which the birth of the backbone is known to be
discontinuous.

Precise numerical simulations \cite{MZKST1999,MZKST_RSA} on the
computational complexity of solving critical 2+p-SAT instances support
the view that the crossover between polynomial and exponential
scalings takes place at $p_0$, the very value of $p$ separating
continuous from discontinuous transitions.  Though investigated
2+p-SAT instances are all critical and the problem itself is
NP-complete for any $p>0$, it is only when the phase transition is
abrupt that hardness shows up (including the fastest known randomized
search algorithms such as walk-sat~\cite{Walksat}).

To understand why search algorithms require polynomial or exponential
computational efforts, statistical studies of the solutions cannot be
sufficient. A full dynamical study of how search procedures operate
has to be carried out. Such studies had already been initiated by
mathematicians in the easy region, where search tree are particularly
simple and almost no backtracking occurs.  Franco and Chao
\cite{FrancoChao} have in particular analyzed the operation of DP
algorithms with different kinds of heuristics and have shown that at small
values of $\alpha$ the typical complexity is linear in $N$.

Recently, the whole range of values of $\alpha$, including the hard
phase, has been investigated, using dynamical statistical mechanics
tools~\cite{CoccoMonasson}. During the search process, the search tree
built by DP grows with time and this growth process can be analyzed
quantitatively. The key idea is that, under the action of DP, 3-SAT
instances are turned into mixed 2+p-SAT instances (some clauses are
simplified into clauses of length two, other are satisfied and
eliminated).  The parameters $p$ and $\alpha$ of the instance under
consideration {\em dynamically} evolve under the action of DP. Their
evolution can be traced back as a trajectory in the phase diagram of
the 2+p-SAT model of figure~\ref{ksat_fig6}. Depending on whether
trajectories cross or not the sat/unsat boundary, easy or hard
resolutions take place, and the location of crossings can be used to
quantitatively predict the scaling of the resolution
times~\cite{CoccoMonasson}.

\newcommand{\spin}{\mathbf{S}}

\section{The traveling salesman problem and the cavity method}
\label{sect_tsp}

In Section~\ref{randgraph}, we derived partition functions using
statistical physics representations based on analytic
continuations. Furthermore, we used the saddle point method on these
partition functions and that allowed us to reproduce a number of exact
results. Then we moved on in Section~\ref{sect_ksat} and applied these
methods to models with quenched disorder. However, because of the
greater complexity of such models, we resorted to an additional tool
of statistical physics: the replica method. Though this kind of
approach is non-rigourous, it is believed that it provides new exact
results for a number of different problems, in particular in
optimization.

The replica method is not the only technical tool that physicists have
developed in the past years. Another approach, called the cavity
method, will be exposed in the present Section.  The cavity approach
gives, at the end of the computation, the sames results as the replica
approach. Yet the assumptions it relies upon turn out to be much more
intuitive and its formalism is closer to a probabilistic theory
formulation. Because of this, it can be used to prove some of the
results derived from statistical mechanics;
see~\cite{Talagrand00,Aldous00} for recent progress in this
direction. In the rest of this section, we show how this cavity method
can be used to ``solve'' a case of the Traveling Salesman Problem
(TSP).

The TSP is probably the world's most studied optimization problem. As
usually formulated for a weighted graph, one considers all Hamiltonian
cycles or ``tours'' (closed circuits visiting each vertex once and
only once) and asks for the shortest one. The total length is given by
the sum of the weights or ``lengths'' of the edges making up the
tour. Since the Hamiltonian cycle problem is NP-complete, certainly
the TSP is very difficult. However, in most cases considered, the
graph is complete (there is an edge for each pair of vertices), so the
difficulty lies in determining the {\it shortest} tour. Without
further restrictions on the nature of the graph, the TSP is {\it
NP}-hard~\cite{GareyJohnson79}.  One speaks of the {\it asymmetric}
TSP when the edges on the graph are oriented, and of the {\it
symmetric} TSP for the usual (unoriented) case. Both types are
frequently used models in scheduling and routing problems, though the
industrial applications tend to move away from the simple formulations
considered in academia.  The symmetric TSPs are further divided into
``metric'' and non-metric according to whether or not the triangle
inequality for the edge lengths is satisfied. The so-called Euclidean
TSP is probably the best known TSP and it is metric; the vertices are
points (cities, or sites) in the plane, and the length of the edge
connecting cities $i$ and $j$ is given by the {\it Euclidean} distance
between $i$ and $j$.  Even within this restricted class of weighted
graphs, the problem of finding the optimum tour remains
NP-hard~\cite{GareyJohnson79}.

The TSP has been at the forefront of many past and recent developments
in complexity. For instance, pretty much all general purpose
algorithmic approaches have been first presented and tested for the
TSP.  This tradition begins back in 1959 when Beardwood et
al.~\cite{BeardwoodHalton59} published tour lengths obtained from
hand-drawn solutions! Later, the idea of optimization by local search
was introduced in the context of the TSP by Lin~\cite{Lin65}, and
simulated annealing~\cite{KirkpatrickGelatt83,Cerny85} was first
tested on TSPs also. The list continues with branch and
bound~\cite{LawlerWood66}, until today's state of the art algorithms
based on cutting planes (branch and cut)\cite{PadbergRinaldi91},
allowing one to solve problems with several thousand
cities~\cite{ApplegateBixby98}. Many physicists have worked on these
kinds of algorithmic questions from a practical point of view; in most
cases their algorithms incorporate concepts such as temperature, mean
field, and renormalization, that are standard in statistical physics,
leading to some of the most effective methods of heuristic
resolution~\cite{JohnsonMcGeoch97}.  It might be argued that these
approaches can also be used to improve the heuristic decision rules at
the heart of exact methods (for instance in branching strategies), but
more work has to be done to determine whether this is indeed the case.

The widespread academic use of the TSP also extends to other 
issues in complexity. For instance,
there has been much recent progress in approximability
of the TSP~\cite{Arora98}. However statistical physics
has nothing to say about worst case behavior;
instead it is relevant for describing
the typical behavior arising in a statistical framework
and tends to focus on self-averaging properties. Thus we are lead
to consider TSPs where the edge lengths
between vertices are chosen randomly according to a given
probability distribution; the corresponding problem is called the
stochastic TSP.

\subsection{The stochastic TSP.}

Statistical physicists as well as probabilists are not interested
per-se in any particular instance of the TSP, rather they seek
``generic'' properties. This might be the typical computational
complexity or the typical length of TSPs with $N$ cities.  It is then
necessary to consider the {\it stochastic} TSP where each instance
(the specification of the weighted graph) is taken at random from an
ensemble of instances; this defines our ``quenched
disorder''. Although one may be interested in many different
ensembles, only a few have been the subject of thorough
investigation. Perhaps the most studied stochastic TSP is the
Euclidean one where the cities are randomly distributed in a given
region of the plane~\cite{BeardwoodHalton59}.  This is a ``random
point'' ensemble. Another ensemble that has been much considered
consists in having the edge lengths all be independent random
variables, corresponding to a ``random distance'' ensemble. (This
terminology is misleading: the problem is not metric as the triangle
inequality is generally not satisfied.) Random distance ensembles have
been considered for both the symmetric~\cite{VannimenusMezard84} and
the asymmetric~\cite{Karp79} TSP.

For any of these ensembles, one can ask for the behavior of the
optimum tour length, or consider properties of the tour itself. Most
work by probabilists has focused on the first aspect
(see~\cite{Steele97} for a review), starting with the seminal work of
Beardwood, Halton, and Hammersley~\cite{BeardwoodHalton59} (hereafter
referred to as BHH). Those authors considered the Euclidean ensemble
where points are randomly (and independently) distributed in a bounded
region $\Omega$ of $d$-dimensional Euclidean space according to the
probability density $\rho({\bf{X}})$. Given a not too singular $\rho$,
BHH proved that the optimum tour length, $L_E$, becomes peaked at
large $N$, and that with probability one as $N \to \infty$
\begin{equation}
\label{eq_BHH}
\frac {L_E}{N^{1-1/d}} \to \beta(d) \int_{\Omega} 
\rho^{1-1/d}({\bf X} ) d{\bf X}
\end{equation} 
Here $\beta$ is a constant, independent of $\rho$, depending only on
the dimension of space.  Some comments are in order.  The first is
that the relative fluctuations of the tour length about its mean tend
to zero as $N \to \infty$, allowing one to meaningfully define a
``typical'' or generic tour length at large $N$.  This fundamental
property was initially proven using sub-additivity properties of the
tour length, but from a more modern perspective, it follows from
considering the passage from $N$ to $N+1$ cities, corresponding to a
martingale process (see~\cite{RheeTalagrand87}).  The second point is
that the $N$ dependence of this typical length is such that the
rescaled length $L_E / N^{1-1/d}$ {\it converges} in probability at
large $N$.  In the language of statistical physics, this quantity is
just the ground state energy density of the system where one increases
the volume linearly with $N$ so that the mean density of points is
$N$-independent. In general such an energy density is expected to be
self-averaging, i.e., have a well defined large $N$ limit, independent
of the sequence of randomly generated samples (with probability one)
as in Eq.(\ref{eq_BHH}). In some problems, the self-averaging property
can be derived, while it will simply be assumed to hold when using the
cavity approach.

Another comment is that given Eq.(\ref{eq_BHH}), the essence of the
problem is the same for any $\rho({\bf X})$; it is thus common
practice to formulate the Euclidean TSP using $N$ points laid down
independently in a unit square (or hypercube if $d >2$), the
distribution being uniform.

There has been much work~\cite{Steele97} on obtaining bounds and
various estimates of the constants $\beta(d)$, but no exact results
are known for $d>1$. However, Rhee~\cite{Rhee92} has proved that
\begin{equation}
\frac {\beta(d)} {\sqrt d} \to \frac {1}{\sqrt {2 e \pi}} ~~~\rm {as} ~~~
d \to \infty
\end{equation}
From the point of view of a statistical physics analysis, the
difficulty in computing $\beta(d)$ arises from the correlations among
the point to point distances.  Indeed, in the Euclidean ensemble,
there are $d N$ random variables associated with the random positions
of the points, and $N (N-1)/2$ distances; these distances are thus
highly redundant (and a fortiori correlated). When these distances are
instead taken to be random and {\it independent}, the ``cavity''
method of statistical physics allows one to perform the calculation of
the corresponding $\beta$. Because of this, we will focus on that
quenched disorder ensemble.

In the ``independent edge-lengths ensemble'' (as opposed to the
independent points ensemble), it is the distances or edge lengths
between points that are independent random variables. Let $d_{ij}$ be
the ``distance'' between points $i$ and $j$ (the problem is not
metric, but we nevertheless follow the standard nomenclature and refer
to $d_{ij}$ as a distance). In the most studied case, $d_{ij}$ is
taken from a uniform distribution in $\lbrack 0, 1 \rbrack$. From a
physicist's perspective, it is natural to stay ``close'' to the
Euclidean random point ensemble~\cite{VannimenusMezard84} by taking
the distribution of $d_{ij}$ to be that of two points randomly
distributed in the unit square (hypercube when $d>2$). The independent
points and independent edge-lengths ensembles then have the same
distribution for individual distances, and in the short distance and
large $N$ limit they also have the same distribution for pairs of
distances. The main difference between the ensembles thus arises when
considering three or more distances; in the Euclidean case, these have
correlations as shown for instance by the triangle inequality.

The minimum tour length in these random edge-lengths models is
expected to be self-averaging; the methods of Rhee and
Talagrand~\cite{RheeTalagrand87} show that the distribution of TSP
tour lengths becomes peaked at large $N$ in this case, but currently
there is no proof of the {\it existence} of a limit as in the
Euclidean case. Nevertheless, this seems to be just a technical
difficulty, and it is expected that the rescaled tour length indeed
has a limit at large $N$; we thus define $\beta(d)$ in analogy to the
expression in Eq.(\ref{eq_BHH}) with the understanding that the
$\beta$s are different in the independent points and independent
edge-lengths ensembles.

\subsection{A statistical physics representation.}

Following the notation of Section~\ref{sect_statisticalPhysics}, we
introduce the generating or partition function
\begin{equation}
\label{eq_Z_tsp}
Z(T) = \sum_{\sigma} \exp({-{\frac{L(\sigma)}{T}}})
\end{equation}
where $\sigma$ is a permutation of the vertices and determines
uniquely a tour. In effect we have identified configurations with
tours, that is with permutations; furthermore, the energy of a
configuration is simply the length of its tour. This construction
amounts to introducing a probability $e^{-L(\sigma)/T}/Z$ for each
tour. When $T = \infty$, all tours are equally probable, while when $T
\to 0$ only the shortest tour(s) survive. As before, $T$ is the
temperature, and the averages $\langle . \rangle_T$ using this
probability distribution are the thermal averages. From them one can
extract most quantities of interest. For instance
\begin{equation}
<L>_T = - \frac{1}{Z} \ \frac{d  Z}{d (1/T)}
\end{equation}
gives the mean tour length at temperature $T$. 
We then have for the TSP tour length: $L_{min} = \lim_{T \to 0} <L>_T$.

The generating function $Z$ requires performing a sum over all
permutations and is a difficult object to treat. To circumvent this
difficulty, a different representation is used. We first introduce
what is called a ``spin'' ${\spin}$, having now $m$-components,
$S^{\alpha}$, $\alpha=1, ..., m$.  These components are real and
satisfy the constraint $\sum_{\alpha} (S^{\alpha})^2 = m$. Such a spin
can be identified with a point on a sphere in $m$-dimensional
Euclidean space. Note that when $m=1$, we recover the kinds of spins
considered in the previous sections. Now for our statistical physics
representation of the TSP, a spin ${\spin}_i$ is associated to each
vertex $V_i$ of the graph, $i=1, ... ,N$.  Define $R_{ij} =
e^{-d_{ij}/T}$ and introduce a new generating function
\begin{equation}
  G(T,m,\omega) = 
\int d{\spin}_1 d{\spin}_2 ... d{\spin}_N \,
\exp(\omega\,\sum_{i<j}R_{ij}\,{\spin}_i\cdot{\spin}_j)
\end{equation}
In this expression, $\cdot$ is the usual scalar product, and
$d{\spin}$ is associated with the uniform measure on the sphere in
dimension $m$. We have normalized it so that $\int d{\spin} = 1$; then
$\int d{\spin} S^{\alpha} S^{\beta} = \delta_{\alpha,\beta}$.  The
claim is now that the initial generating function $Z$ is equivalent to
using an analytic continuation of $G$ in $m$:
\begin{equation}
\label{eq_G_tsp}
\lim_{\scriptstyle{m\to 0}\atop\scriptstyle{\omega\to\infty}}
\frac{G-1}{m\omega^N} \equiv \sum_{\sigma} \exp({-{\frac{L(\sigma)}{T}}})
\end{equation}
Comparing to the Potts model of Section~\ref{randgraph}, we see that
$m$ is analogous to the Potts parameter $q$: the partition function is
defined for integer values of the parameter, and then has to be
analytically continued to real values.

The derivation of equation~(\ref{eq_G_tsp}) is based on 
showing the equality of both sides when performing a
power series in $1/T$. First expand the exponential in the integral:
\begin{equation}
\label{eq_G_tsp_expand}
G = \int d{\spin}_1 d{\spin}_2 ... d{\spin}_N \,
\left[ 1 + \omega\,\sum_{i<j}R_{ij}
({\spin_i}\cdot
{\spin}_j) + \frac{\omega^2}{2!}\,\cdots \right]
\end{equation}
Now integrate term by term; each resulting contribution can be
associated with a subgraph (but where edges can appear multiple times)
whose weight is given in terms of its edges and its cycles.  (Note
that each vertex must be covered an even number of times because the
integrand is even under $\spin_i \to -\spin_i$.) Each edge $E_{ij}$
appearing in the subgraph contributes a multiplicative factor $R_{ij}$
to its total weight. A further factor comes from the loops (cycles) of
the subgraph. It is not difficult to see that each such loop leads to
a factor $m$ in the total weight because of the integration over the
$m$-dimensional spins. Thus as $m \to 0$ only subgraphs having a
single loop survive in $G$ and then vertices cannot belong to more
than two edges.  Finally, when $\omega \to \infty$, the loops with the
most vertices dominate, leading to tours. Thus if we first take $m \to
0$ and {\it then} $\omega \to \infty$, the expansion of $G-1$ reduces
to a sum over all the tours of the graph. Furthermore, the weight of
each tour is proportional to the product of the $R_{ij}$ belonging to
the tour, so that one recovers the total weight $m \omega^N \exp
(-L/T)$ where $L$ is the tour length.  In conclusion,
Eq.~(\ref{eq_G_tsp}) is justified to all orders in $1/T$, and thus for
any finite $N$ it holds as an identity.

Whether one uses $Z$ or $G-1$ does not matter as they differ only by
an irrelevant multiplicative factor (we assume $m$ and $1/\omega$
infinitesimal).  From $G-1$, one can compute the optimum tour and not
just the optimum tour length; indeed, at finite temperature, the
probability that a tour contains the edge $E_{ij}$ is given by the
mean occupation of that edge. Defining $n_{ij}=1$ if the edge is used
by the tour and $n_{ij}=0$ otherwise, the probability of occupation is
\begin{equation}
\label{eq_occupation}
\langle n_{ij} \rangle_T = \omega R_{ij} 
\langle {\spin}_i\cdot{\spin}_j \rangle_T 
\end{equation}
where from now on $\langle . \rangle_T$ means thermal average using
{\it either} $Z$ or $G-1$; the one that is used should be clear from
the observable considered. Now if we take in Eq.~\ref{eq_occupation}
the limit $T \to 0$, we find those edges that are occupied and thus
the optimal tour (assuming it is unique).  Note also that
Eq.~(\ref{eq_occupation}) has a simple justification: $\langle
{\spin}_i\cdot{\spin}_j \rangle_T$ has a numerator whose expansion
gives $m \omega^{N-1} / R_{ij}$ times the weighted sum over all tours
containing the edge $ij$, while the denominator is $m \omega^N$ times
the weighted sum over all tours. The identity
Eq.~(\ref{eq_occupation}) then follows immediately.

\subsection{The cavity equations.}
\begin{figure}
\begin{center}
\resizebox{0.4\linewidth}{!}{\includegraphics{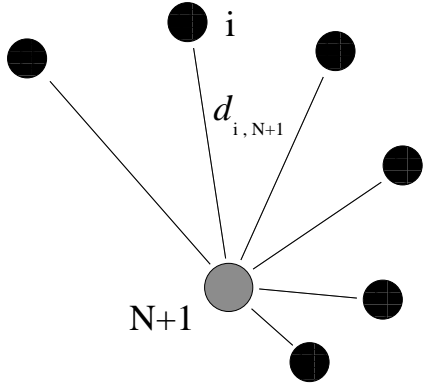}}
\end{center}
\caption{($N+1$)th spin and its ordered neighbors.}
\label{fig_cavity}
\end{figure}

The partition function $G-1$ gives the ``statistical physics'' of the
TSP for any given graph. Using this formalizm to determine
analytically the optimum tour in a general case seems an impossible
task. Nevertheless, $G$ is a good starting point for following the
passage from $N$ to $N+1$ vertices as in a martingale process, and the
derivation of a recursion in $N$ is the heart of the cavity
method. The term cavity comes from the fact that the system at $N+1$
is compared to the one at $N$ by removing the ($N+1$)th spin, thereby
creating a cavity.  In figure~\ref{fig_cavity}, we have represented in
counter-clockwise order the nearest, next-nearest, etc...  neighbors
of site $N+1$ which is at the center of the cavity.
Because the total number of spins will be sometimes $N$ and sometimes
$N+1$, we indicate the number via a subscript on $G$. Thus for
instance $G_N-1$ is to be used when considering quantities for the
system with $N$ spins.  Now for every quantity associated with the
system having $N+1$ spins, if we integrate explicitly over spin $N+1$,
we are left with quantities defined in the system having only $N$
spins. Consider for instance $G_{N+1}-1$ itself.  When expanding the
exponentials depending on ${\spin}_{N+1}$, we obtain: (i) terms linear
in ${\spin}_{N+1}$ that integrate to zero; (ii) terms quadratic in
${\spin}_{N+1}$ that upon integration give products
${\spin}_i\cdot{\spin}_j$; (iii) higher powers in ${\spin}_{N+1}$ that
do not contribute as $m \to 0$. A simple calculation leads to the
identity
\begin{equation}
\label{eq_G_recurrence}
\frac{G_{N+1} - 1}{G_N - 1} = \omega^2 \sum_{1\le j<k\le N}
R_{j,N+1} R_{k,N+1} 
\langle {\spin}_j \cdot {\spin}_k \rangle_T^{'} =  \frac{Z_{N+1}}{Z_N}
\end{equation}
where $\langle . \rangle_T^{'}$ is a ``cavity average'', to be taken
in the system having only the first $N$ spins, spin $N+1$ being
absent. Note that $Z_N$ and $Z_{N+1}$ are the partition functions of
Eq.~\ref{eq_Z_tsp} when there are $N$ and $N+1$ vertices; also, it is
easy to see that one need not restrict the sum to $j \ne k$ because
the term $j=k$ vanishes as $m \to 0$.

Straightforward calculations in this same spirit lead to relations
between thermal expectation values using $N+1$ spins and those using
$N$ spins. For instance
\begin{equation}
\label{eq_recur1}
\langle {\spin}_{N+1} \rangle_T ~ (G_{N+1} - 1 ) = \sum_{j=1}^N 
\omega R_{j,N+1}\langle {\spin}_j \rangle_T^{'} ~ (G_N - 1 )
\end{equation}
Similarly, one has for the two-spin average:
\begin{equation}
\label{eq_recur2}
\langle {\spin}_{N+1} \cdot {\spin}_i \rangle_T ~ (G_{N+1} - 1 ) = 
\sum_{j\ne i} \omega R_{j,N+1} ~ 
\langle {\spin}_i \cdot {\spin}_j \rangle_T^{'} ~ (G_N - 1 )
\end{equation}
More generally, the numerator in any observable depending on spin
$N+1$ has a simple expression in terms of the numerators of
observables in the absence of that spin. Furthermore, one can use
Eq.~\ref{eq_G_recurrence} to eliminate all reference to $G_{N}$ and
$G_{N+1}$ in these relations.  The conclusion is that if we know how
to compute the properties of systems with $N$ spins, we can then
deduce those of systems with $N+1$ spins; the cavity method is thus a
recursion on $N$ for all the properties of such a system.

\subsection{The factorization approximation.}
Unfortunately, these recursion equations cannot be solved, but let us
approximate them by neglecting certain correlations.  Clearly,
${\spin}_{N+1}$ is strongly correlated with its nearest neighbors
because the corresponding $R$s are important. More generally, two
spins whose joining edge length is short (are near neighbors) will be
strongly correlated because short tours will often occupy that
edge. Thus we must and will take into account the correlations between
${\spin}_{N+1}$ and its near neighbors. However, we will neglect here
the correlations among these neighbors themselves, so that in the
absence of ${\spin}_{N+1}$, their joint probability distribution
factorizes, so that in particular
\begin{equation}
\label{eq_tsp_factorization}
\langle {\spin}_i \cdot {\spin}_j \rangle_T^{'}  = 
\langle {\spin}_i \rangle_T^{'}  \cdot \langle {\spin}_j \rangle_T^{'}
\end{equation}
This property implies that replica symmetry is not broken, and this is
indeed believed to be the case for the TSP. Factorization makes the
cavity approach particularly tractable, as we shall soon see.  (In
systems where replica symmetry {\it is} broken, it is necessary to
find ways to parametrize these correlations; this is quite complex and
not well resolved, even within the statistical physics approach.)

A second point concerns the meaning of $\langle {\spin}_{N+1}
\rangle_T$. $G_{N+1}$ is rotationally symmetric; there is no preferred
direction, so the thermal average of any spin vanishes. Note however
that we have seen a similar situation before in the context of the
Ising model (c.f. Section~\ref{sect_statisticalPhysics}). Here as
before, the interactions tend to align the spins. Thus, when the
temperature is low enough, we expect to have a spontaneous
magnetization when $N \to \infty$. To make this more explicit, we can
introduce a small magnetic field, i.e., an interaction term of the
type $-{\mathbf{h}} \cdot {\spin}_i$ for each spin; we then take the
limit $N \to \infty$ and only after take $\mathbf{h} \to 0$. This
magnetic field breaks the rotational symmetry, and so the system has a
preferred direction, even after the field has been removed. By
convention, we shall take this direction to be along the first axis.

Given these two remarks, we can use the exact equations
(\ref{eq_recur1}) and (\ref{eq_recur2}) to obtain the cavity equations
assuming factorization. Denoting by $S^1$ the component along the
first axis of $\spin$, one has
\begin{equation}
\label{eq_cav1}
\langle S_{N+1}^1\rangle_T = \frac{\sum_{j=1}^N 
R_{j,N+1}\langle S_j^1\rangle_T^{'}}
{\omega\,\sum_{1\le j<k\le N}
R_{j,N+1} R_{k,N+1} ~ \langle S_j^1\rangle_T^{'} ~ \,\langle 
S_k^1\rangle_T^{'}}
\end{equation}
Similarly, one has for the two-spin average 
(see Eq.~\ref{eq_occupation}):
\begin{equation}
\label{eq_cav2}
\langle n_{i,N+1} \rangle_T = R_{i,N+1}
\langle S_i^1\rangle_T^{'} ~ 
\frac{\sum_{j\ne i}R_{j,N+1} ~ \langle S_j^1\rangle_T^{'}}
{\sum_{1\le j<k\le N}
R_{j,N+1} R_{k,N+1} 
\langle S_j^1\rangle_T^{'} ~ \,\langle S_k^1\rangle_T^{'}}
\end{equation}
These are the standard cavity recurrence equations, first derived by
M\'ezard and Parisi~\cite{MezardParisi86}.  We also note that in this
factorization approximation, one has $\langle {\spin}_{N+1} \cdot
{\spin}_i \rangle_T =
\langle S_{N+1}^1  \rangle_T  \langle S_i^1  \rangle_T^{'}$

\subsection{The $N \to \infty$ and $T \to 0$ limits.}
The last step of the cavity method is to assume that the recurrence
equations, when considered in the disorder ensemble, give rise to a
stationary stochastic process when $N \to \infty$. Consider for
instance the individual magnetizations $\langle {\spin}_{i}\rangle_T$;
they are random variables because the $d_{ij}$ themselves are. If we
want them to have a limiting distribution at large $N$, (i.e., in
physical terms, to have a thermodynamic limit), we have to rescale the
$d_{ij}$ by $N^{1/d}$ or equivalently set $T={\tilde T} N^{-1/d}$ with
${\tilde T}$ fixed.  (Note that in the case of the Euclidean TSP, the
rescaling of lengths can be interpreted as taking the limit $N \to
\infty$ while keeping the density of points fixed, that is by
increasing the size of the volume $\Omega$ linearly with $N$.) The
important point is that the ``environment'' seen by the spins must
have limiting statistical properties as $N \to \infty$, and this
translates to having $N$-independent statistics for the distances of a
spin to its near neighbors. Then it is assumed that the probability
density of the $\langle S^1_i\rangle_T$ converges to a limiting
distribution $P_{\infty}$ when $N \to \infty$.  The cavity method is
thus a kind of {\it bootstrap} approach where $P_{\infty}$ is assumed
to exist and it is determined by its stationarity property under the
cavity recurrence.

That such a stationary limit exists can be motivated by the large $N$
behavior of the tour length in the stochastic TSP. In fact, it is
expected that all quantities associated with any fixed number of edges
will converge in the thermodynamic limit, so it should be possible to
look at $2$, $3$, or $k$ edge constructs.  At present though, because
of the technical difficulty, only the single edge computations have
been carried out.  Fortunately, that is enough for getting the value
of $\beta$, and allows one to obtain the so called link-length
distribution, i.e., the distribution of the edge lengths appearing in
the optimal tours.

Equation~(\ref{eq_cav1}) with the condition of stationarity of the
stochastic process leads to a complicated implicit equation for
$P_{\infty}$. Fortunately, in the zero temperature limit (which is
where we recover the usual stochastic TSP), the recurrence relations
are much simpler. Following Krauth and M\'ezard~\cite{KrauthMezard89},
one defines $\phi_i$ for any vertex $i$ ($i=1, ..., N$) via:
\begin{equation}
\langle S_i^1 \rangle^{'} = 
\frac{\exp(\phi_i / {\tilde T} )}{\omega^{1/2}}
\end{equation}
One also defines $\phi_{N+1}$ analogously 
using $\langle S_{N+1}^1 \rangle$. Now re-order
the indices of the first $N$ vertices so that
\begin{equation}
N^{1/d} d_{1,N+1} - \phi_1 \le 
N^{1/d} d_{2,N+1} - \phi_2 \le ... \le 
N^{1/d} d_{N,N+1} - \phi_N 
\end{equation}
Then the zero-temperature limit
of Eq.~\ref{eq_cav1} leads to
\begin{equation}
\label{eq_tsp_phi}
\phi_{N+1} = d_{2,N+1} N^{1/d} - \phi_2 
\end{equation}
while Eq.~\ref{eq_cav2} shows that the optimum tour uses the edges
connecting $N+1$ to vertices $1$ and $2$, i.e., $n_{1,N+1} = n_{2,N+1}
= 1$, all others are equal to zero.

If we have a stationary stochastic process, Eq.~(\ref{eq_tsp_phi})
leads to a self-consistent equation for the probability density $P$ of
the $\phi$s.  We also see that the random variables $\chi_i = N^{1/d}
d_{i,N+1} - \phi_i$ ($i=1, ..., N$) play a fundamental role. By
hypothesis, they are uncorrelated: the $d_{i,N+1}$ because we are
dealing with the independent edge-lengths ensemble, and the $\phi_i$
because we have explicitly neglected the correlations between the
spins in the absence of ${\spin}_{N+1}$. Denote by $\Pi(\chi)$ the
probability density of these random variables; $\Pi(\chi)$ is uniquely
determined in terms of $P$, assuming the distribution of $d_{i,N+1}$
given. From here on, take for simplicity these edge lengths to be
uniformily distributed in $\lbrack 0,1 \rbrack$.  (This corresponds to
the $1$-dimensional case $d=1$; we refer the reader
to~\cite{KrauthMezard89} for more general distributions.)  The
relation between $\Pi$ and $P$ then becomes
\begin{equation}
\Pi (x) = {\frac{1}{N}} \int_0^N P(l-\chi) dl
\end{equation}
Now a self-consistent equation for $P$ is obtained by using the fact
that $\phi_{N+1}$ is the second smallest of the $N$ different $\chi$s:
\begin{equation}
P(\phi) = N (N-1) \Pi(\phi) ( \int_{-\infty}^{+\infty}
\Pi(u) du ) ( \int_{\phi}^{+\infty} \Pi(u) du )^{N-2}
\end{equation}
In the large $N$ limit, this integral non-linear implicit equation
simplifies to
\begin{equation}
P(\phi) = {\frac{d G(\phi)}{d \phi}} \, G(\phi) \, e^{- G(\phi)} 
\, \, \, \, {\rm where} \, \, \, \,
G(\phi) = \int_0^{+\infty} u P(u - \phi) du 
\end{equation}
Plugging the expression for $P$ into this last equation leads to
\begin{equation}
G(\phi) = \int_{-\phi}^{+\infty} \lbrack 1 + G(t) \rbrack ~ e^{- G(t) } ~ dt
\end{equation}
This cannot be solved analytically, but can easily be treated
numerically, and one can obtain machine precision results for $G$ and
thus $P$ without too much effort.

Assuming $G$ and $P$ have been computed, one can find in a similar way
the distribution of $d_{1,N+1}$ and $d_{2,N+1}$. For instance, the
distribution of the rescaled distance $N d_{1,N+1} = \tilde l_1$ is
given by
\begin{equation}
P_1(\tilde l_1) = \int_{-\infty}^{+\infty} P(\tilde l_1 - \chi) 
e^{- G(\chi)} d \chi
\end{equation}
This, along with the analogous distribution for $d_{2,N+1}$, gives the
distribution of edge lengths in the optimum tour, and thus also the
mean tour length, i.e., when $d=1$, the value of $\beta$. Krauth and
M\'ezard~\cite{KrauthMezard89} showed that this constant could be
written in terms of $G$ alone,
\begin{equation}
\beta = {\frac {1}{2}} \int_{-\infty}^{+\infty} 
G(t) \, \lbrack 1 + G(t) \rbrack \, e^{- G(t)} \, dt
\end{equation}
and they found $\beta = 2.041$\ldots (Note that when $d=1$, as
suggested by Eq.~(\ref{eq_BHH}), the tour length becomes independent
of $N$. This can be understood qualitatively by observing that each
vertex can connect to one of its near neighbors that is at a distance
$O(1/N)$.)

\subsection{``Exact'' solution in the independent edge-lengths ensemble.}
As described, the cavity method involves an uncontrollable
approximation associated with ignoring certain correlations. It is
natural to ask whether those correlations might in fact be absent in
certain ensembles. A simple case is when the graph considered is a
Cayley tree with the root (corresponding to vertex $N+1$) removed.
Then the different neighbors of ${\spin}_{N+1}$ are uncoupled and have
no correlations at all. Unfortunately, this type of graph will not do
for the TSP as it has no Hamiltonian cycles, but it can do for other
problems close to the TSP such as the minimum matching problem.

So let us consider instead the structure of {\it independent
edge-lengths} graphs. {\it Locally} their properties ressemble those
of Cayley trees, so that with some luck the previous reasoning can
hold for these types of graphs as $N \to \infty$.  Although the
correlations that were neglected in the cavity calculation will always
be present at finite $N$ in the independent edge-lengths model, they
have every reason to go to zero as $N \to \infty$. The justification
is that the close neighbors of vertex $N+1$ are ``infinitely'' far
from one-another when $N \to \infty$. In the language of tours (rather
than spins), this means that the probability for the tour to have an
edge connecting two of the finite order neighbors of vertex $N+1$
should go to zero at large $N$. Clearly this is not the case in the
Euclidean stochastic TSP because of the triangle inequality: the
neighbor of a neighbor is itself a neighbor. But in the independent
edge-length model, the neighbors represented in
figure~\ref{fig_cavity} are ``far away'' from one-another with a
probability tending towards $1$ as $N \to \infty$. This kind of random
``geometry'' is then expected to lead to uncorrelated spins among the
finite order neighbors of ${\spin}_{N+1}$ and so the cavity
calculation may become exact as $N \to \infty$.

Although it is not clear yet that the correlations go away as $N \to
\infty$ in the independent edge-lengths ensemble, the reasoning above
is supported by extensive simulational results. In these kinds of
tests, one generates weighted graphs in the ensemble of interest,
determines the optimum tour for different sizes $N$, and then
estimates the statistical properties in the large $N$ limit. All such
simulational studies to date have confirmed the validity of the cavity
method. Both the assumptions of no replica symmetry
breaking~\cite{PercusMartin99} and the predictions for $\beta$ and
$P(d_{1,N+1})$ have been
validated~\cite{KrauthMezard89,Johnson_HK,PercusMartin99} in that
way. Although these tests have limited precision in the context of the
TSP, more stringent tests~\cite{BrunettiKrauth91,HoudayerBoutet98}
have been performed on matching problems. For instance, using the
cavity and replica methods, M\'ezard and Parisi
predicted~\cite{MezardParisi86} that the length of a minimum matching
of $N$ points would have the large $N$ limit $\pi^2 / 12$ when the
$d_{ij}$ are uniformily distributed in $\lbrack 0, 1 \rbrack$. The
numerical simulations confirm this value at the level of $0.05\%$.

The consensus is thus that the cavity method gives {\it exact} results
at large $N$ for all independent edge-lengths disorder ensembles. But
for the physicist, this is not the only interest of the cavity method:
even as an approximation, it is useful for understanding the effects
of quenched disorder. For instance, one can ask~\cite{KrauthMezard89}
how bad is the factorization approximation when applied to the
Euclidean TSP in $d=2$. For that, we compare Krauth and M\'ezard's
cavity prediction $\beta(2) = 0.7251...$ to the best estimate from
numerical simulations~\cite{PercusMartin96,Johnson_HK} $0.7120 \pm
0.0004$. We see that in fact the prediction is quantitatively good,
and it turns out that this approximation becomes even better as the
dimension of space $d$ is increased.

\bigskip
\subsection{Remarks on the cavity approach and replica symmetry breaking.}
In some respects, the cavity method is complementary to the replica
method, but both become unwieldy when replica symmetry is broken. In
the case of the TSP, it turns out that only the cavity method has
allowed a complete solution, but that model has no replica symmetry
breaking. When replica symmetry breaking does arise, the situation is
far more complex, and to date only models defined on graphs with
infinite connectivity have been solved exactly (though not
rigorously).  Nevertheless, recent progress~\cite{MezardParisi00} in
using the cavity method may soon lead to ``exact'' solutions of other
models such as K-SAT in spite of the presence of replica symmetry
breaking.

\section{Related topics and conclusion.}

\subsection{Other optimization problems investigated in physics.}
\label{othertopics}

This article has focused on presenting statistical physics tools in
the context of a few well-known problems.  But many other random
combinatorial problems have been considered by physicists, often using
nearly identical techniques to the ones we have presented. For the
reader interested in having a more complete view of such work, we give
here a partial list of problems and pointers to the litterature.

\subsubsection*{Graph bipartitioning.}

Given a graph $G$, partition its $N$ vertices into two sets of equal
size. The cost of the partition is the number of edges connecting
vertices in different sets. The graph bipartitioning (or graph
bisection) problem consists in finding the minimum cost partition.

This problem is readily reformulated in the physics language of spins:
to each vertex $i$ attach a spin $S_i$ and set it to $+1$ if the
vertex is assigned to the first set and $-1$ if it is assigned to the
second set.  Calling $G_{ij}$ the adjacency matrix of the graph $G$,
the number of edges ``crossing'' the partition can be identified with
an energy:
\begin{equation}
E= \frac 12 \sum _{i<j} G_{ij}   \left( 1 - S_i S_j \right)
\qquad .
\end{equation} 
Since the partition is assumed balanced, the global magnetization
$M=\sum _i S_i$ is constrained to be zero. In physics studies,
researchers enforce this constraint in a soft way by adding $\lambda
M^2 /2$ to the energy $E$, where $\lambda$ is a positive parameter.
As a result, spins interact through effective couplings $J_{ij} =
(G_{ij}-\lambda)/2$ that can be positive or negative.  The
corresponding energy function is then seen to be a spin glass
Hamiltonian, similar to the Sherrington--Kirpatrick model exposed in
Section~\ref{refsk}.  The first authors to notice this identification
were Fu and Anderson\cite{FuAnderson86,Fischer}. They then applied the
Parisi solution of the Sherrington--Kirpatrick model to give the large
$N$ value of the minimum cost partition when $G$ has connectivities
growing linearly with $N$. These results generalize to weighted graphs
straightforwardly.

\subsubsection*{Weighted minimum bipartite matching.}
Let $I$ and $J$ be two sets containing $N$ points each.  We assume
given an $N\times N$ matrix of ``distances'' $d_{ij}$ defined for each
pair $i\in I, j\in J$. For any complete matching (a one-to-one map or
a pairing between $I$ and $J$, more commonly known as a bipartite
matching), its cost is defined as the sum of the distances between
paired points. In the {\it minimum} weighted bipartite matching
problem one is to find the complete matching of lowest
cost. Naturally, one can consider a stochastic version where the
entries of the distance matrix are independent random variables, drawn
from a probability distribution $p(d)$. This problem is close in its
technical aspects to the stochastic TSP, and like the non-bipartite
case it has been ``solved'' both via the replica and the cavity
methods~\cite{MezardParisi86,MezardParisi87}. In the special case
where $p(d)$ is the uniform distribution in $[0,1]$, M\'ezard and
Parisi hav computed the large $N$ limit of the typical cost to be
$\pi^2 / 6$. In fact, in a real {\it tour de force}, they also
obtained the form of the $1/N$ correction to this limit. More
recently, Parisi considered the special case $p(d) = \exp(-d)$ and
conjectured~\cite{Parisi98} that for any $N$ the mean minimum cost is
given by $\sum_{k=1, ..., N} 1/k^2$. All current evidence, both
numerical and analytical for small $N$ values~\cite{Dotsenko00},
indicates that this formula at finite $N$ could be exact.

\subsubsection*{Number partitioning.} 

This problem can be motivated by the need to divide an estate between
two inheritors in a fair way. It is usually formulated as follows. Let
$\{ x_1, x_2, ..., x_N \}$ be $N$ real numbers in $[0,1]$ and consider
a partition of the $x_i$ into two (un-balanced) sets. The
``unfairness'' of a partition is the sum of the $x$'s in the first set
minus the sum of the $x$'s in the second.  The number partitioning
problem consists in determining the partition that minimizes the
absolute value of the unfairness.  When the $x_i$ are independent
random numbers, it is possible to derive some statistical properties
of the minimum.  We refer the reader to Mertens' detailed review in
the present issue~\cite{Mertensrev} of his recent work.

\subsubsection*{Vertex cover} 

Very recently, A. Hartmann and M. Weigt studied the minimum size of
vertex coverings of random graphs.  Phase transitions take place,
accompanied by drastic changes of the computational complexity of
finding optimal vertex coverings using branch--and--bound
algorithms. See the article in the present volume \cite{HartWeigt}.

\subsubsection*{Neural Networks.}
To a large extent, learning and generalization properties of formal
neural networks are optimization problems. These properties have been
the subject of intense studies by statistical physicists in the last
fifteen years.  A quite complete review of these works and results are
exposed in the article by A. Engel in this volume~\cite{Engel}.

\subsection{Further statistical properties.}

Statistical physics concepts and techniques are powerful tools to
investigate the properties of ground states, that is the solutions of
combinatorial optimization problems. So far, we have concentrated on
the large size (large ``$N$'') limit of these problems, but one can
also consider finite $N$. In addition, it may be of interest to know
the properties of the near-optimum solutions.

\subsubsection*{Finite-size corrections and scaling.}

Mean-field models can be solved through saddle-point calculations in
the infinite size limit only. Clearly, optimization problems usually
deal with a {\it finite} number of variables. It is therefore crucial
to achieve a quantitative understanding of the finite size corrections
to be expected, e.g., on the ground state energy.

Far from phase transitions, corrections to the saddle-point value can
usually be computed in a systematic way using perturbation theory. An
example of such a calculation to determine finite-size corrections has
been mentioned previously (see the bipartite matching problem
discussed in Section~\ref{othertopics}).  For any quantity or
``observable'' associated with the optimum solution of a problem, one
can ask how its disorder-average depends on the system
size. Similarly, fluctuations, which disappear in the infinite volume
limit, generally matter for finite sizes.  Both effects are well-known
in the statistical physics of systems {\em without disordered
interactions} and have been the subject of many theoretical
studies\cite{cardy,privman}.

Close to transition points, the handling of finite-size corrections is
much more involved. Few results are avalaible for disordered systems
\cite{parisisk}. Generally speaking, the transition region is characterized
by a window, the width of which scales as some negative power of the
system size, shrinking to zero in the infinite size limit. We have
already discussed the critical scaling properties of some systems in
Sections~\ref{critexpo1} and
\ref{critexpo2}. No similar theoretical study of critical exponents
has been performed so far for complex optimization problems, e.g. K-SAT;
only numerical data or bounds on the exponents are currently available.

\subsubsection*{Finite-dimensional energy landscapes and robustness.}

Realistic physical systems and certain optimization problems such as
the Traveling Salesman Problem live in a finite-dimensional world.
Thus, although we considered in Section~\ref{randgraph} a percolation
model on a random graph, the physics of the problem is usually modeled
using a lattice in two or three-dimensional space, edges joining
vertices only if they are close in Euclidean space.  Models based on
random graphs are considered to describe physical systems only when
the dimension goes to $\infty$.

Finite-dimensionality may have dramatic consequences on some
properties of the models; for instance it is known that the critical
exponents depend on the dimension of the embedding space. More
crucially, in low dimensions, the correct order parameter could be
quite different from what it is in infinite dimension.  This issue is
particularly acute in the physics community in the case of spin
glasses: so far, no consensus has been reached concerning the correct
description of these systems in dimension $3$. Two main theories
exist:

\begin{itemize}

\item{\em Parisi's hierarchical picture.} This sophisticated theory
comes from extending mean-field theory to finite dimensional spaces.
It states that low lying configurations, {\em i.e.} having an energy
slightly larger than the ground state, may be very far away, in the
configuration space, from the ground state. These excited
configurations are organized in a complex hierarchical fashion, in
fact an ultrametric structure.

\item{\em The droplet picture.} Conversely, the droplet picture is based
on simple scaling arguments inspired from ferromagnetic systems and
claims that low-lying configurations stand close to the ground state.
Higher and higher energy excitations will be obtained when flipping
more and more spins from the ground state.

\end{itemize}
A detailed presentation of the theories can be found
in~\cite{Young98}.  Knowing which picture is actually correct could
have deep consequences for dynamical issues (see the next paragraph),
and also for the robustness of the ground state. For instance, it can
be important from a practical point of view to know how much a
perturbation or modification of the energy function affects the ground
state properties. Consider in particular the problem of image
reconstruction. Can a small change in the data modify macroscopically
the reconstructed image?  Within the droplet picture, the answer would
be generally no, while Parisi's theory would support the view that
disordered systems often have non-robust ground states.

\subsection{Perspectives.}
The study of the statistical properties of disordered systems has
witnessed major advances in the last two decades, but the most recent
trend has been towards trans-disciplinary applications. Although it is
difficult to guess what new directions will emerge, there has been a
clear and growing interest in using statistical physics tools for
investigating problems at the heart of computer science.  In this
review, we illustrated this for decision and optimization problems,
but many other problems should follow. Looking at the most recent
work, we see emerging efforts to extend these methods to understand
the statistical properties of the corresponding algorithms, be-they
exact or heuristic. Let us first sketch these issues and then mention
some further possible directions.

\subsubsection*{Typical case computational complexity.}
The notion of {\it typical case} computational complexity is
appealing, and statistical physics tools may help one understand how
that kind of classification of decision problems may be reached.  But
clearly the methods needed to do so go much beyond what we have
presented: partition functions and analogous tools describe the
solutions of a problem, not how long it can take to find
them. Nevertheless, as we mentioned in Section~\ref{subsect_ccd} in
the context of the Davis-Putnam tree search, physical arguments can
shed new light on how algorithms such as branch and bound behave near
a phase transition. Thus these methods may tell us what is the typical
computational complexity of an instance chosen at random in an
ensemble, given a particular tree search algorithm. Extending this
classification to obtain an algorithm-independent definition of
typical case computational complexity may follow, but so far it
remains largely open.

\subsubsection*{Long time (stationary) limit of stochastic search algorithms.}
Consider heuristic algorithms that are based on stochastic
search. Examples are simulated annealing, G-Walk, or deterministic
limits of these such as local search. These kinds of algorithms define
random walks, i.e., stochastic dynamics on a discrete space of
solutions (boolean assignments for K-SAT, tours for the TSP, etc...)
and these dynamics are ``local'': just a few variables are changed at
each time step. Assume for simplicity that the initial position of the
walk is chosen at random.  At long times, the search settles in a
steady state where the distribution of energies becomes stationary,
that is time-independent.  (The energy at any given time is a random
variable, depending on the starting point of the search and also on
all the steps of the walk up to that time.  The energy thus has a
distribution when considering all initial positions and all possible
walks.) An obvious question is whether this distribution becomes
peaked in the large size limit. Indeed, in most cases, one can show
that the energy of a {\it random} solution is self-averaging; note
that this corresponds simply to the self-averaging property of the
thermodynamic energy at infinite temperature. In fact, for the
problems we have focused upon, the energy is expected to be
self-averaging {\it at all temperatures}. By a not so bold
extrapolation, one may conjecture that any local stochastic search
algorithm leads to self-averaging energies in the long time limit.
(Naturally, we also have to assume that the algorithms do not have too
much memory; using a simulated annealing with temperatures changing
periodically in time will not do!)  There is numerical
evidence~\cite{SchreiberMartin99} in favor of this conjecture, and it
may be possible to use statistical physics methods to prove it in some
limiting cases. One can also ask what is the limiting shape of the
{\it distribution} of energies. This is a difficult question, but it
may be easier in this context than when considering the optimum.
\subsubsection*{Dynamics of stochastic search algorithms.}
Is the self-averaging behavior just mentioned restricted to long times?
Since the initial energies are those of random solutions
and are thus self-averaging, it is quite natural to generalize
the conjecture to all times: ``the energy at any given time
of a local stochastic search algorithm is self-averaging''.

Quite a bit of intuition about this issue can be obtained by
considering what happens by analogy with a physical system relaxing
towards equilibrium.  The main characteristic of the dynamics in a
physical system is the property called detailed balance; this
condition puts very stringent restrictions on the transition
probabilities. But within this specific framework, there has been much
progress recently in describing the time dependence of the dynamical
process. In particular, the conjecture introduced above is confirmed
in the context of mean field $p$-spin glass models. The exact solution
of these models has led to new results on entropy production while the
phenomenon of ``ageing'' has been explained theoretically. Clearly an
important goal is to extend these results to arbitrary stochastic
dynamics without the hypothesis of detailed balance. But perhaps one
of the most remarquable results coming from these studies (see for
instance the contribution of Bouchaud et al. in\cite{Young98}) is a
relation between the relaxation during these dynamics and the effects
of a perturbation: the prediction, called the generalized
fluctuation-dissipation relation, seems numerically to be quite
general and it would be of major interest to test it in the context of
more general stochastic dynamics.

\subsubsection*{Further directions.}
We will be brief and just give a list of what we consider to be
promising topics. First, just as the notion of computational
complexity has to be generalized to a typical case description, the
analogous generalization of approximability is of interest.  In its
stochastic or typical extension, an algorithm provides an $\epsilon$
typical case approximation to a problem if with probability tending
towards $1$ in the large size limit, its output is within $\epsilon$
of the actual solution. Naturally results that hold in the worse case
also hold stochastically, but one may expect new properties to hold in
this generalized framework.  Second, there has been an upsurge of
interest in physics for combinatoric problems, using techniques from
field theory and quantum gravity. The problems range from coloring
graphs to enumerating meanders. Although the initial problem has no
disorder, the approaches use identities relating systems with disorder
to systems without disorder that are as yet still in the conjectoral
stage.  Third, is there a relation between replica symmetry breaking
and typical case complexity? Forth, will the statistical physics
approaches in artificial neural networks and learning lead to new
developments in artificial intelligence? Fifth, an active subject of
study in decision science concerns ``belief propagation'' algorithms
which are extensions of the cavity method. Can these extensions lead
to better understanding of physical systems, and inversely, will the
use of physics concepts such as temperature, mean field, scaling, and
universality continue to lead to improved algorithms in practice?

\paragraph*{Acknowledgements ---}
O.C.M. acknowledges support from the Institut Universitaire de
France. We thank C. Kenyon and P. Flajolet for comments
and encouragements.

\newpage

\appendix
\section{Answers to Exercises}

\subsection{Exercise 1: System with two spins and statistical independence.}

The partition function (\ref{foncpart}) at temperature $T=1/\beta $ reads
\begin{eqnarray}
Z (T) &=& \sum _{\sigma _1 , \sigma _2 = \pm 1} 
\exp \left( -\frac 1T \; E(\sigma _1 , \sigma _2 ) \right) \nonumber \\
&=& \sum _{\sigma _1 , \sigma _2 = \pm 1} 
\exp \left( \beta \sigma_1 \sigma_2 ) \right) \nonumber \\
&=& 4 \cosh \beta \qquad .
\end{eqnarray}
The magnetization $m(T)$ and the average value of the energy $\langle E
\rangle _T $ can be computed from the knowledge of $Z$, see
(\ref{ret5}). One obtains
\begin{equation}
m(T) = \langle \sigma _1 \rangle _T= 0 \quad ,
\end{equation}
and
\begin{equation}
\langle E \rangle _T =  -  \tanh (\beta )  \quad .
\end{equation}
The magnetization vanishes since any configuration $\{ \sigma _1 ,
\sigma _2 \}$ has the same statistical weight as its opposite,
$\{ -\sigma _1 ,- \sigma _2 \}$.
 
These calculations can be repeated for the second choice of the energy
function, $E(\sigma_1,\sigma_2) = - \sigma_1 - \sigma_2$, with the
following results:
\begin{eqnarray}
Z (T) &=& (2 \; \cosh \beta ) ^2 \nonumber \\
m(T) &=& \tanh \beta \nonumber \\
\langle E \rangle _T &=&  - 2\; \tanh (\beta )  \qquad .
\end{eqnarray}
We see that the partition function is the square or the 
single spin partition function.
The magnetization and the energy (once
divided by the number of spins) are equal to the ones of a single
spin, see expression (\ref{mag1}). 

This coincidence is a direct consequence of the additivity property
of the energy. More precisely, whenever the energy of a system can be
written as the sum of two (or more) energies of disjoint subsystems, {\em
i.e.}, involving disjoint configuration variables, the
partition function is simply the product of the subsystems partition
functions. Such disjoint subsystems do not interact and are
statistically independent.

\subsection{Exercise 2: Zero temperature energy and entropy.}

Let us suppose that the configurations ${ C}$ form a
discrete set. Let us call $E_0$ the smallest energy and ${ N}_0$
the number of configurations having this energy. Similarly we call
$E_1$ the immediately higher value of energy, with degeneracy ${
N}_1$. This process can be repeated for more and more excited
energies. At the end, configurations are sorted according to their
energies with $E_0 < E_1 < E_2 < \ldots$. 

From the definition (\ref{foncpart}) of the partition
function, we write 
\begin{eqnarray}
Z &=& \sum _{j \ge 0} { N}_j \; e^{- \beta \; E_j } \nonumber \\
&=& e^{- \beta \; E_0 } \; \bigg( { N}_0 + { N}_1 
\; e^{- \beta \; G_1 } + { N}_2 
\; e^{- \beta \; G_2 } + \ldots \bigg) \qquad ,
\end{eqnarray}
where $G_j = E_j - E_0$ is the gap between the $j^{th}$ excited energy
and the minimal one. By construction, all gaps $G_j$ are strictly
positive ($j\ge 1$). Thus, in the small temperature (large $\beta$)
limit, we obtain
\begin{equation}
Z(T) = { N}_0 \; e^{- \beta \; E_0 } \bigg( 1 + O\bigg( e^{- \beta \;
G_1 } \bigg) \bigg)
\quad ,
\end{equation}
from which we deduce the free-energy,
\begin{equation}
F(T) = -T \ln Z(T) = E_0 - T \ln { N}_0  + O\bigg(\frac 1\beta \;
 e^{- \beta \; G_1 } \bigg) 
\quad .
\end{equation}
From the definition of entropy (\ref{entro}), it appears that the zero
temperature entropy $\langle S \rangle _{T=0}$
is simply  the logarithm of the number of absolute
minima of the energy function $E({ C})$.

\subsection{Exercise 3: Spins on the complete graph in the 
presence of a field.}

The calculations are immediate from (\ref{isingf}). The only
difference is that, in the presence of a small but non zero field $h$, the two 
minima of the free-energy shown on figure~\ref{isingelib} are now at
two different heights. One of the two minima (with the opposite sign of
$h$) is exponentially suppressed with respect to the other.

\subsection{Exercise 4: Quenched average.}

Using the results of Exercise 2, we write the partition function,
magnetization and the average value of the energy,
\begin{eqnarray}
Z (T,J) &=& 4 \; \cosh ( \beta J) \nonumber \\
m(T,J) &=& 0 \nonumber \\
\langle E \rangle _T (J) &=&  -  J\; \tanh (\beta J)  \qquad .
\end{eqnarray}
All these statistical quantities depend on the quenched coupling $J$. 

We now average over the coupling $J$, with distribution $\rho (J)$ on
the support $[J_-; J_+]$. We obtain for the quenched  average 
magnetization and energy, 
\begin{eqnarray}
\overline{ m(T)} &=& 0 \nonumber \\
\overline{\langle E \rangle _T} &=&  - \int _{J_-} ^{J_+} dJ \; \rho
(J) \; J \; \tanh (\beta J)  \qquad .
\label{avth}
\end{eqnarray}
In the zero temperature limit, the spins align (respectively
anti-align) onto each other if the coupling $J$ is positive
(resp. negative). The resulting ground state energy equals $|J|$. 
Averaging over the quenched coupling, we obtain
\begin{equation}
\overline{\langle E \rangle _{T=0}} =  - \int _{J_-} ^{J_+} dJ \; \rho
(J)  |J|
\quad .
\end{equation}

\subsection{Exercise 5: Frustrated triangle of spins.}

Both energies are even functions of the spins; the magnetization is
thus always equal to zero.

We first consider the energy function 
\begin{equation}
E(\sigma_1,\sigma_2 , \sigma _3) = - \sigma_1 \sigma_2 - \sigma_1 \sigma_3 -
 \sigma_2 \sigma_3 \quad .
\end{equation}
The partition function and the average value of the energy read respectively,
\begin{eqnarray}
Z (T) &=& 2\; e^{3 \beta} + 6 \; e^{-\beta } \nonumber \\
\langle E \rangle _T &=& \frac{ -3 +3  e^{-4 \beta }}
{1+ 3 e^{-4 \beta }} \qquad .
\end{eqnarray}
In the zero temperature limit, the ground state energy and entropy are
given by
\begin{eqnarray}
\langle E \rangle _{T=0} &=& -3 \nonumber \\
\langle S \rangle _{T=0} &=& \ln 2\qquad .
\end{eqnarray}
There are indeed two configurations with minimal energy; all their spins are
aligned in the same direction.

We now consider the energy function 
\begin{equation}
E(\sigma_1,\sigma_2, \sigma _3) = - \sigma_1 \sigma_2 - \sigma_1 \sigma_3 +
 \sigma_2 \sigma_3 \quad .
\end{equation}
The partition function and the average value of the energy now read 
respectively,
\begin{eqnarray}
Z (T) &=& 6\; e^{\beta} + 2 \; e^{-3 \beta } \nonumber \\
\langle E \rangle _T &=& \frac{ -3 +3  e^{-4 \beta }}
{3+  e^{-4 \beta }}  \qquad .
\end{eqnarray}
In the zero temperature limit, the ground state energy and entropy are
given by
\begin{eqnarray}
\langle E \rangle _{T=0} &=& -1 \nonumber \\
\langle E \rangle _{T=0} &=& \ln 6\qquad .
\end{eqnarray}
As a result of frustration, the ground state energy is higher than in
the previous case, as well as the number of ground
states. Note also that the gap between the lowest and second lowest
energy levels has become smaller.

\subsection{Exercise 6: Partition function of the
 Sherrington-Kirkpatrick model.}

The partition function of the Sherrington-Kirkpatrick (SK) model
reads
\begin{equation}
Z( { J} )  = \sum _{\sigma _i = \pm 1} \exp \left( \frac{\beta}
{\sqrt N} \sum _{i< j} J_{ij} \sigma _i \sigma _j \right)  \quad ,
\end{equation}
where the quenched couplings ${ J} = \{ J_{ij} , 1\le i < j \le N \}$ 
are randomly drawn from the Gaussian distribution
\begin{equation}
{ P} ( { J} )  = \prod _{1\le i < j \le N } \frac 1{\sqrt{2
\pi}} \; \exp \left(- \frac 12  J_{ij} ^2 \right) \quad .
\end{equation}
To compute the average value of the partition function, we first
average the couplings out and only then calculate the sum over the spins
\begin{eqnarray}
\overline{Z( { J} ) } &=& \int d{ J} \; { P} ( { J} )
\; Z( { J} )  \nonumber \\
&=&  \sum _{\sigma _i = \pm 1} \exp \left( \frac{\beta ^2}{2N} \sum _{i<
 j} (  \sigma _i \sigma _j) ^2 \right)  \nonumber \\
&=&  2^N \; \exp \left( \frac{\beta ^2} 4 (N-1) \right) 
\qquad ,
\label{skz1}
\end{eqnarray}
We now calculate the second moment of the partition function by
rewriting the squared sum as the product of two independent sums, see
Exercise~1, 
\begin{eqnarray}
\overline{Z( { J} )^2 } &=& \int d{ J} \; { P} ( { J} )
\; Z( { J} ) ^2   \nonumber \\
&=& \int d{ J} \; { P} ( { J} ) \;
\sum _{\sigma _i = \pm 1} \sum _{\tau _i = \pm 1}  \exp \left( \frac{\beta}
{\sqrt N} \sum _{i< j} J_{ij} ( \sigma _i \sigma _j 
+ \tau _i \tau _j ) \right) \nonumber \\
&=&  \sum _{\sigma _i = \pm 1} \sum _{\tau _i = \pm 1} 
\exp \left( \frac{\beta ^2}{2N} \sum _{i<
 j} (  \sigma _i \sigma _j + \tau _i \tau _j) ^2 \right)  \nonumber \\
&=& \bigg( \overline{Z( { J} ) } \bigg) ^2 \; Y 
\qquad ,
\label{skz21}
\end{eqnarray}
where $Y$ equals
\begin{eqnarray}
Y &=& \frac 1{4^N} \;  \sum _{\sigma _i = \pm 1} \sum _{\tau _i = \pm
1}  \exp \left( \frac{\beta ^2}
{N} \sum _{i< j} \sigma _i \sigma _j 
\tau _i \tau _j  \right) \nonumber \\
&=& \frac 1{4^N} \;  \exp \left( - \frac{\beta ^2}2 \right)
\sum _{\sigma _i = \pm 1} \sum _{\tau _i = \pm
1}  \exp \left( \frac{\beta ^2} {2 N}  \left[ \sum _{i} \sigma _i \tau
_i \right] ^2  \right)
\quad . \label{Y}
\end{eqnarray}
The calculation proceeds as in the case of the spin model on the
complete graph, see section~\ref{dep}. 
We define for each configuration ${ C} = \{\sigma
_i , \tau _i \}$ of the $2N$ spins, the overlap function
\begin{equation}
q({ C} ) = \frac 1N \sum _{i=1} ^N  \sigma _i \tau
_i \qquad . \label{overlup}
\end{equation}
The effective energy function appearing in the last term of the pseudo
partition function $Y$ (\ref{Y}) depends on the configuration through
$q({ C})$ only. Following the steps of section~\ref{dep}, 
a saddle-point calculation leads to the asymptotic behaviour of $Y$, 
\begin{equation}
Y = \exp \left( - N \beta ^2 \phi ^* + o(N) \right) \qquad ,
\end{equation}
where $\phi ^*$ is the minimum over $q$ of the ``free-energy''
functional $\hat f (q)$ defined in (\ref{isingf}) with $T^2$ instead
of $T$. The results of section~\ref{dep} teach us that there is a
``critical'' temperature $T_c =1$ such that $\phi^*=0$ for
temperatures above $T_c$ and $\phi^* < 0$ when $T<T_c$. 

Above $T_c$, the partition function does not fluctuate too much around
the average value $\overline{Z( { J} ) }$; the partition function
is itself self-averaging and the free-energy per spin simply equals
$f(T)= -T \ln 2 - 1 / (4T)$, see the paper by M. Talagrand in the same
volume. At low temperatures, below $T_c$, the second moment of $Z(
{ J} )$ is exponentially larger than the squared average; there
are huge fluctuations and the partition function is not
self-averaging. It is therefore much more complicated to calculate the
value of the free-energy.

\subsection{Exercise 7: A toy replica calculation.}

We want to compute the series expansion of $ln (1+x)$ starting from
the identity (for small real $n$)
\begin{equation} \label{ne0}
(1+x)^n = 1 + n \; \ln (1+x) + O( n^2) \qquad ,
\end{equation}
and the series expansion of $(1+x)^n$ for integer n. To do so,
we use Newton's binomial formula
\begin{equation}
(1+x)^n = \sum _{k=0} ^n \frac{n!}{k! (n-k)!} \; x^k \qquad ,
\label{ne1}
\end{equation}
valid for positive integers $n$. $n$ play two roles in formula
(\ref{ne1}). First, it is the upper bound of the sum over
$k$. Secondly, $n$ appears in the combinatorial factor in the
sum. Factorials may be continued analytically to real values of $n$
using Euler's Gamma function. As $\Gamma (z)$ has poles at negative
integer values of the argument $z$, we may extend the sum in expression
(\ref{ne1}) to integer values of $k$ larger than $n$ without changing the
final result,
\begin{equation}
(1+x)^n = \sum _{k=0} ^\infty \frac{n!}{k! (n-k)!} \; x^k \qquad .
\label{ne2}
\end{equation}
Let us focus now on the combinatorial factor
\begin{equation}
C(n,k) = \frac{n!}{k! (n-k)!} = \frac{n(n-1) (n-2) 
\ldots (n-k+1)} {k!}  \qquad  .
\label{ne3}
\end{equation}
For $k=0$, we have $C(n,0)=1$ for all $n$. 
When $k\ge 1$, the r.h.s. of (\ref{ne3}) is a polynomial of $n$ and can be
immediately continued to real $n$. In the small $n$ limit, we obtain 
\begin{equation}
C(n,k) = n \; \frac{ (-1) (-2) \ldots (-k) }{k!} + o(n) = n \; \frac{
(-1)^{k-1}}{k} + o(n) \qquad (k \ge 1)\ . 
\label{ne4}
\end{equation}
Finally, we write the small $n$ continuation of equation (\ref{ne1})
as
\begin{equation}
(1+x)^n = 1 + n \; \sum _{k=1} ^\infty \frac{ (-1)^{k-1}}{k}\; x^k 
+o (n) \qquad .
\label{ne5}
\end{equation}
Comparing equation (\ref{ne0}) and (\ref{ne5}), we obtain the correct
result
\begin{equation}
\ln (1+x) = \sum _{k=1} ^\infty \frac{ (-1)^{k-1}}{k}\; x^k 
 \qquad .
\label{ne6}
\end{equation}
The above calculation is a simple application of the replica
trick. Obviously, the calculation of the free-energy of disordered
models, e.g. the K-Satisfiability or the TSP models, are much more 
involved from a technical point of view.


\newpage

\addcontentsline{toc}{chapter}{\protect\bibname}
\begin{thebibliography}{10}
\expandafter\ifx\csname url\endcsname\relax
  \def\url#1{\texttt{#1}}\fi
\expandafter\ifx\csname urlprefix\endcsname\relax\def\urlprefix{URL }\fi

\bibitem{MezardParisiVirasoro}
M.~M\'ezard, G.~Parisi, M.~A. Virasoro (Eds.), Spin Glass Theory and Beyond,
  World Scientific, Singapore, 1987.

\bibitem{Young98}
A.~P. Young (Ed.), Spin Glasses and Random Fields, World Scientific, Singapore,
  1998.

\bibitem{HartWeigt}
A.~Hartmann, M.~Weigt, Statistical mechanics perspective on the phase
  transition in vertex covering of finite-connectivity random graphs, in the
  present issue of Theoretical Computer Science .

\bibitem{Mertensrev}
S.~Mertens, A physicist's approach to number partitioning, in the present issue
  of Theoretical Computer Science .

\bibitem{Engel}
A.~Engel, Complexity of learning in artificial neural networks, in the present
  issue of Theoretical Computer Science .

\bibitem{baxter}
R.~Baxter, Exactly solved models in Statistical Mechanics, Academic Press, San
  Diego, 1982.

\bibitem{Rieger98}
H.~Rieger, Frustrated systems: Ground state properties via combinatorial
  optimization, in: J.~Kertesz, I.~Kondor (Eds.), Advances in Computer
  Simulation, Vol. 501 of Lecture Notes in Physics, Springer-Verlag,
  Heidelberg, 1998.

\bibitem{Youngrev}
P.~Young, Informatics - 10 years back, 10 years ahead, celebration of the 10th
  aniversary of Schloss Dagstuhl .

\bibitem{Reif}
R.~Reif, Fundamentals of Statistical and Thermal Physics, McGraw-Hill,
  New-York, 1965.

\bibitem{Ma}
S.~Ma, Statistical Mechanics, World Scientific, Singapore, 1985.

\bibitem{Huang}
K.~Huang, Statistical Mechanics, Wiley, New-York, 1967.

\bibitem{Saenger84}
W.~Saenger, Principle of Nucleic Acid Structure, Springer-Verlag, New-York,
  1984.

\bibitem{Strick98}
T.~Strick, J.~Allemand, D.~Bensimon, R.~Lavery, V.~Croquette, Phase coexistence
  in a single {DNA} molecule, Physica A 263 (1998) 392.

\bibitem{Steele97}
J.~M. Steele, Probability Theory and Combinatorial Optimization, SIAM,
  Philadelphia, 1997.

\bibitem{GareyJohnson79}
M.~R. Garey, D.~S. Johnson, Computers and Intractability: A Guide to the Theory
  of {NP}-Completeness, Freeman, New York, 1979.

\bibitem{PapadimitriouSteiglitz}
C.~H. Papadimitriou, K.~Steiglitz, Combinatorial Optimization: Algorithms and
  Complexity, Prentice Hall, Englewood Cliffs, NJ, 1982.

\bibitem{Bollo}
B.~Bollob\`as, Random Graphs, Academic Press, New-York, 1985.

\bibitem{Potts52}
R.~Potts, Proc. Camb. Phil. Soc. 48 (1952) 106.

\bibitem{Kasteleyn}
P.~Kasteleyn, C.~Fortuin, J. Phys. Soc. Japan Suppl. 26 (1969) 1114.

\bibitem{Wu82}
F.~Wu, The potts model, Rev. Mod. Phys. 54 (1982) 235.

\bibitem{Morgante}
A.~Morgante, Large deviations in random graphs, Tech. rep., Laboratoire de
  Physique Theorique de l'ENS, rapport de stage (1998).

\bibitem{MonassonZecchina96}
R.~Monasson, R.~Zecchina, Entropy of the {K}-satisfiability problem, Phys. Rev.
  Lett. 76 (1996) 3881--3885.

\bibitem{MonassonZecchina97}
R.~Monasson, R.~Zecchina, Statistical mechanics of the random {K-S}at problem,
  Phys. Rev. E 56 (1997) 1357--1361.

\bibitem{MonassonZecchina98}
R.~Monasson, R.~Zecchina, Tricritical points in random combinatorics: the
  $(2+p)$-{SAT} case, J. Phys. A 31 (1998) 9209--9217.

\bibitem{MZKST1999}
R.~Monasson, R.~Zecchina, S.~Kirkpatrick, B.~Selman, L.~Troyansky,
  Computational complexity from 'characteristic' phase transitions, Nature 400
  (1999) 133--137.

\bibitem{MZKST_RSA}
R.~Monasson, R.~Zecchina, S.~Kirkpatrick, B.~Selman, L.~Troyansky, 2+p-sat:
  Relation of typical-case complexity to the nature of the phase transition,
  Random Structures and Algorithms 3 (1999) 414.

\bibitem{Monasson_diluiti}
R.~Monasson, Optimization problems and replica symmetry breaking in finite
  connectivity spin glasses, J.~Phys. A 31 (1998) 513.

\bibitem{Variational}
G.~Biroli, R.~Monasson, M.~Weigt, A variational description of the ground state
  structure in random satisfiability problems, Euro. Phys. J. B 14 (2000) 551.

\bibitem{Papadimitriou}
C.~H. Papadimitriou, Computational Complexity, Addison--Wesley, 1994.

\bibitem{Cook}
S.~Cook, The complexity of theorem--proving procedures, in: Proc. 3rd Ann. ACM
  Symp. on Theory of Computing, Assoc. Comput. Mach., New York, 1971, p. 151.

\bibitem{Hogg}
T.~H. Eds., B.~A. Huberman, C.~Williams, Issue 1-2, special issue on phase
  transitions, Artificial Intelligence 81.

\bibitem{hardness}
D.~Mitchell, B.~Selman, H.~Levesque, Hard and easy distributions of sat
  problems, in: Proc.\ of Am. Assoc. for Artif. Intell. AAAI-92, 1992, pp.
  456--465.

\bibitem{2SAT_1}
A.~Goerdt, A threshold for unsatisfiability, J. Comput. System Sci. 53 (1996)
  469.

\bibitem{2SAT_2}
V.~Chv\`atal, B.~Reed, Mick gets some (the odds are on his side), in: Proc.
  33rd IEEE Symp. on Foundations of Computer Science, 1992, p. 620.

\bibitem{KirkpatrickSelman}
S.~Kirkpatrick, B.~Selman, Critical behaviour in the satisfiability of random
  boolean expressions, Science 264 (1994) 1297.

\bibitem{Self_averaging}
A.~Broder, A.~Frieze, E.~Upfal, On the satisfiability and maximum
  satisfiability of random 3-cnf formulas, Proc. 4th Annual ACM-SIAM Symp. on
  Discrete Algorithms  (1993) 322.

\bibitem{Rigorous_entropy}
O.~Dubois Private communication.

\bibitem{RicciWeigtZecchina}
F.~Ricci-Tersenghi, M.~Weigt, R.~Zecchina, The simplest k-satisfiability model,
  Phys. Rev. E 63 (2001) 026702, arXiv:cond-mat/0011181.

\bibitem{BanavarSherrington87}
J.~R. Banavar, D.~Sherrington, N.~Sourlas, Graph bipartitioning and statistical
  mechanics, J. Phys. A Lett. 20 (1987) L1--L8.

\bibitem{MezardParisi87c}
M.~M{\'e}zard, G.~Parisi, Mean-field theory of randomly frustrated systems with
  finite connectivity, Europhys. Lett. 3~(10) (1987) 1067.

\bibitem{VianaBray85}
L.~Viana, A.~J. Bray, Phase diagrams for dilute spin-glasses, J. Phys. C 18
  (1985) 3037--3051.

\bibitem{KanterSompolinsky87}
I.~Kanter, H.~Sompolinsky, Mean-field theory of spin-glasses with finite
  coordination number, Phys. Rev. Lett. 58 (1987) 164.

\bibitem{DeDomGold}
C.~D. Dominicis, Y.~Goldschmidt, Replica symmetry breaking in finite
  connectivity systems: a large connectivity expansion at finite and zero
  temperature, J.~Phys. A 22 (1989) L775.

\bibitem{MottDeDom}
P.~Mottishaw, C.~D. Dominicis, On the stability of randomly frustrated systems
  with finite connectivity, J.~Phys. A 20 (1987) L375.

\bibitem{DeDomMott}
C.~D. Dominicis, P.~Mottishaw, Replica symmetry breaking in weak connectivity
  systems, J.~Phys. A 20 (1987) L1267.

\bibitem{GoldLai}
Y.~Goldschmidt, P.~Lai, The finite connectivity spin glass: investigation of
  replica symmetry breaking of the ground state, J.~Phys. A 23 (1990) L775.

\bibitem{ScalingWindow}
B.~Bollob\`as, C.~Borgs, J.~Chayes, J.~H. Kim, D.~Wilson, The scaling window of
  the 2-sat transition, Random Structures and Algorithms 2000, in press.

\bibitem{Wilson}
D.~Wilson, The empirical values of the critical k-sat exponents are wrong 2000
  preprint arXiv:math/0005136.

\bibitem{LeoneRicciZecchina}
M.~Leone, F.~Ricci-Tersenghi, R.~Zecchina, Phase coexistence and finite size
  scaling in random combinatorial problems, J.Phys.A, in press (2001).

\bibitem{Walksat}
B.~Selman, H.~Kautz, B.~Cohen, Local search strategies for satisfiability
  testing, in: Proceedings of DIMACS, 1993, p. 661.

\bibitem{FrancoChao}
J.~F. M-T.~Chao, Probabilistic analysis of a generalization of the unit-clause
  literal selection heuristics for the k-satisfiability, Information Science 51
  (1990) 289--314.

\bibitem{CoccoMonasson}
S.~Cocco, R.~Monasson, Trajectories in phase diagrams, growth processes and
  computational complexity: how search algorithms solve the 3-satisfiability
  problem,, Phys. Rev. Lett. 86 (2001) 1654, arXiv:cond-mat/0009410.

\bibitem{Talagrand00}
M.~Talagrand, Rigorous low temperature results for the p-spin mean field spin
  glass model, Probability Theory and Related Fields 117 (2000) 303--360.

\bibitem{Aldous00}
D.~J. Aldous, The zeta(2) limit in the random assignment problem
  Math.PR/0010063.

\bibitem{BeardwoodHalton59}
J.~Beardwood, J.~H. Halton, J.~M. Hammersley, The shortest path through many
  points, Proc. Camb. Phil. Soc. 55 (1959) 299--327.

\bibitem{Lin65}
S.~Lin, Computer solutions of the traveling salesman problem, Bell System
  Technical Journal 44 (1965) 2245--2269.

\bibitem{KirkpatrickGelatt83}
S.~Kirkpatrick, C.~D. {Gelatt Jr.}, M.~P. Vecchi, Optimization by simulated
  annealing, Science 220 (1983) 671--680.

\bibitem{Cerny85}
V.~{\v C}erny, Thermodynamical approach to the traveling salesman problem: An
  efficient simulation algorithm, J.~Optimization Theory Appl. 45 (1985) 41.

\bibitem{LawlerWood66}
E.~L. Lawler, D.~E. Wood, Branch-and-bound methods: A survey, Operations Res.
  14, No. 4 (1966) 699.

\bibitem{PadbergRinaldi91}
M.~W. Padberg, G.~Rinaldi, A branch and cut algorithm for the resolution of
  large-scale symmetric traveling salesman problems, SIAM Review 33 (1991) 60.

\bibitem{ApplegateBixby98}
D.~Applegate, R.~Bixby, V.~Chvátal, W.~Cook, On the solution of traveling
  salesman problems, Documenta Mathematica, J.D.M. ICM III (1998) 645--656.

\bibitem{JohnsonMcGeoch97}
D.~S. Johnson, L.~A. McGeoch, The traveling salesman problem: A case study in
  local optimization, in: E.~H.~L. Aarts, J.~K. Lenstra (Eds.), Local Search in
  Combinatorial Optimization, J. Wiley \& and Sons, New York, 1997, pp.
  215--310.

\bibitem{Arora98}
S.~Arora, Polynomial time approximation schemes for {E}uclidean traveling
  salesman and other geometric problems, Journal of the ACM 45 (1998) 753--782.

\bibitem{VannimenusMezard84}
J.~Vannimenus, M.~M{\'e}zard, On the statistical mechanics of optimization
  problems of the travelling salesman type, J. Physique Lett. 45 (1984)
  L1145--L1153.

\bibitem{Karp79}
R.~M. Karp, A patching algorithm for the nonsymmetric travelling salesman
  problem, SIAM J. Comput. 8 (1979) 561--573.

\bibitem{RheeTalagrand87}
W.~T. Rhee, M.~Talagrand, Martingale inequalities and {NP}-complete problems,
  Mathematics of Operation Research 12~(1) (1987) 177--181.

\bibitem{Rhee92}
W.~T. Rhee, On the travelling salesperson problem in many dimensions, Random
  Structures and Algorithms 3~(3) (1992) 227--233.

\bibitem{MezardParisi86}
M.~M{\'e}zard, G.~Parisi, Mean-field equations for the matching and the
  travelling salesman problem, Europhys. Lett. 2 (1986) 913--918.

\bibitem{KrauthMezard89}
W.~Krauth, M.~M\'ezard, The cavity method and the travelling-salesman problem,
  Europhys. Lett. 8 (1989) 213--218.

\bibitem{PercusMartin99}
A.~G. Percus, O.~C. Martin, The stochastic traveling salesman problem: Finite
  size scaling and the cavity prediction, J. Stat. Phys. 94~(5/6) (1999)
  739--758.

\bibitem{Johnson_HK}
D.~S. Johnson, L.~A. McGeoch, E.~E. Rothberg, Asymptotic experimental analysis
  for the {H}eld-{K}arp traveling salesman bound, in: 7th Annual {ACM-SIAM}
  Symposium on Discrete Algorithms, Atlanta, GA, 1996, pp. 341--350.

\bibitem{BrunettiKrauth91}
R.~Brunetti, W.~Krauth, M.~M{\'e}zard, G.~Parisi, Extensive numerical
  simulations of weighted matchings: Total length and distribution of links in
  the optimal solution, Europhys. Lett. 14 (1991) 295--301.

\bibitem{HoudayerBoutet98}
J.~Houdayer, J.~H. {Boutet de Monvel}, O.~C. Martin, Comparing mean field and
  {E}uclidean matching problems, Eur. Phys. Jour. B 6 (1998) 383--393.

\bibitem{PercusMartin96}
A.~G. Percus, O.~C. Martin, Finite size and dimensional dependence in the
  {E}uclidean traveling salesman problem, Phys. Rev. Lett. 76 (1996)
  1188--1191.

\bibitem{MezardParisi00}
M.~M{\'e}zard, G.~Parisi, The {B}ethe lattice spin glass revisited
  Cond-mat/0009418.

\bibitem{FuAnderson86}
Y.~Fu, P.~W. Anderson, Application of statistical mechanics to {NP}-complete
  problems in combinatorial optimization, J. Phys. A 19 (1986) 1605--1620.

\bibitem{Fischer}
K.~Fischer, J.~Hertz, Spin glasses, Cambridge University Press, Cambridge,
  1991.

\bibitem{MezardParisi87}
M.~M{\'e}zard, G.~Parisi, On the solution of the random link matching problems,
  J. Physique 48 (1987) 1451--1459.

\bibitem{Parisi98}
G.~Parisi, A conjecture on random bipartite matching, cond-mat/9801176 (1998).

\bibitem{Dotsenko00}
V.~Dotsenko, Exact solution of the random bipartite matching model, J. Phys. A
  33 (2000) 2015, cond-mat/9911477.

\bibitem{cardy}
J.~Cardy, Finite size scaling, vol.2 of Current Physics Sources and Comments,
  North-Holland, Amsterdam, 1988.

\bibitem{privman}
V.~Privman, Finite size scaling and numerical simulations of statistical
  systems, World Scientific, Singapore, 1988.

\bibitem{parisisk}
G.~Parisi, F.~Ritort, F.~Slanina, Several results on the finite-size
  corrections in the {S}herrington-{K}irkpatrick spin glass model., J. Phys. A
  26 (1993) 3775.

\bibitem{SchreiberMartin99}
G.~R. Schreiber, O.~C. Martin, Cut size statistics of graph bisection
  heuristics, SIAM Journal on Optimization 10(1) (1999) 231--251.

\end{thebibliography}
\end{document}